\documentclass[fleqn,usenatbib]{mnras}

\usepackage{newtxtext,newtxmath}

\usepackage[T1]{fontenc}
\usepackage{ae,aecompl}

\usepackage{graphicx}	
\usepackage{amsmath}	
\usepackage{amssymb}	
\usepackage{multirow}

\newcommand{\bm}[1]{\boldsymbol{#1}}
\newcommand{\numberOfSimulations}{{$74$}}
\newcommand{\sm}{{\ast}}
\newcommand{\modified}{black}

\makeatletter
\DeclareFontFamily{OMX}{MnSymbolE}{}
\DeclareSymbolFont{MnLargeSymbols}{OMX}{MnSymbolE}{m}{n}
\SetSymbolFont{MnLargeSymbols}{bold}{OMX}{MnSymbolE}{b}{n}
\DeclareFontShape{OMX}{MnSymbolE}{m}{n}{
    <-6>  MnSymbolE5
   <6-7>  MnSymbolE6
   <7-8>  MnSymbolE7
   <8-9>  MnSymbolE8
   <9-10> MnSymbolE9
  <10-12> MnSymbolE10
  <12->   MnSymbolE12
}{}
\DeclareFontShape{OMX}{MnSymbolE}{b}{n}{
    <-6>  MnSymbolE-Bold5
   <6-7>  MnSymbolE-Bold6
   <7-8>  MnSymbolE-Bold7
   <8-9>  MnSymbolE-Bold8
   <9-10> MnSymbolE-Bold9
  <10-12> MnSymbolE-Bold10
  <12->   MnSymbolE-Bold12
}{}
\let\llangle\@undefined
\let\rrangle\@undefined
\DeclareMathDelimiter{\llangle}{\mathopen}%
                     {MnLargeSymbols}{'164}{MnLargeSymbols}{'164}
\DeclareMathDelimiter{\rrangle}{\mathclose}%
                     {MnLargeSymbols}{'171}{MnLargeSymbols}{'171}
\makeatother

\title[Nonlinear outcome of gravitational instability]{Nonlinear outcome of gravitational instability in an irradiated protoplanetary disc}

\author[S. Hirose and J. Shi]{
Shigenobu Hirose,$^{1}$\thanks{E-mail: hirose.shigenobu@gmail.com (SH)}
and Ji-Ming Shi$^{2}$
\\
$^{1}$Department of Mathematical Science and Advanced Technology, JAMSTEC, Yokohama 236-0001, Japan\\
$^{2}$Department of Astrophysical Sciences, Princeton University, 4 Ivy Ln, Princeton, NJ 08544\\
}

\date{Accepted XXX. Received YYY; in original form ZZZ}

\pubyear{2017}

\begin{document}
\label{firstpage}
\pagerange{\pageref{firstpage}--\pageref{lastpage}}
\maketitle

\begin{abstract}
{\color{\modified}
  Using local three dimensional radiation hydrodynamics simulations, the nonlinear outcome of gravitational instability in an irradiated protoplanetary disc is investigated in a parameter space of the surface density $\Sigma$ and the radius $r$.
  Starting from laminar flow, axisymmetric self-gravitating density waves grow first. Their self-gravitating degree becomes larger when $\Sigma$ is larger or the cooling time is shorter at larger radii. The density waves eventually collapse owing to non-axisymmetric instability, which results in either fragmentation or gravito-turbulence after a transient phase.
  The boundaries between the two are found at $r \sim 75$ AU as well as at the $\Sigma$ that corresponds to the initial Toomre's parameter of $\sim 0.2$. The former boundary corresponds to the radius where the cooling time becomes short, approximating unity. Even when gravito-turbulence is established around the boundary radius, such a short cooling time inevitably makes the fluctuation of $\Sigma$ large enough to trigger fragmentation. On the other hand, when $\Sigma$ is beyond the latter boundary (i.e. the initial Toomre's parameter is less than $\sim 0.2$), the initial laminar flow is so unstable against self-gravity that it evolves into fragmentation regardless of the radius or, equivalently, the cooling time. Runaway collapse follows fragmentation when the mass concentration at the centre of a bound object is high enough that the temperature exceeds the H$_2$ dissociation temperature.


}
\end{abstract}




\section{Introduction}

Recent observations, including those by the Atacama Large Millimeter/submillimeter Array (ALMA), have revealed massive protostellar/protoplanetary discs in young stellar class 0 and class I systems \citep[e.g.][]{Andrews:2013,Najita:2014,Perez:2016}. In such massive discs, self-gravity is a very important and relevant aspect of physics. Specifically, when the condition
\begin{align}
  Q \equiv \frac{c_\text{s}\kappa}{\pi G\Sigma} < 1 \label{eq:toomre_condition}
\end{align}
is met, the disc is subject to gravitational instability \citep[GI;][]{Toomre:1964}. Here, $Q$ is called Toomre's parameter whilst $c_\text{s}$ is the sound speed, $\kappa$ is the epicyclic frequency, and $\Sigma$ is the surface density. In Keplerian discs, $\kappa$ is equal to the orbital frequency $\Omega$.

The nonlinear development of GI generally leads to formation of spiral density waves; especially when they are tightly wound, they may be described as so-called gravito-turbulence in the local approximation \citep[see][for a recent comprehensive review on GI in protoplanetary discs]{Kratter:2016}. The shear stress associated with the spiral density waves radially transfers angular momentum, which evolves the radial structure of the disc. Another outcome of GI is fragmentation, or formation of self-gravitationally bound objects, which may eventually become companion stars, brown dwarfs, or gas giant planets. Thus, the nonlinear outcome of GI largely affects the growth and evolution of the disc, but in different forms depending on whether formation of spiral density waves (gravito-turbulence) or fragmentation occur. Therefore, what determines the nonlinear outcome of GI is of great interest, and thus has been widely explored by numerical hydrodynamics simulations.

In the framework of the shearing box, \citet{Gammie:2001} first revealed the importance of the cooling time. He showed that fragmentation occurred when cooling was fast enough, where $\beta \equiv t_\text{cool}\Omega < 3$, when the cooling time $t_\text{cool}$ was assumed constant everywhere for simplicity.
Since then, the fragmentation condition in terms of $\beta$ has been the main focus of interest. It has been extensively studied using various types of numerical methods and cooling prescriptions, both in local and global simulations \citep[e.g.][]{Johnson:2003,Stamatellos:2009,Cossins:2010,Baehr:2015,Rios:2016}, and especially for protoplanetary discs by many authors mostly motivated by the formation of gas giants via GI \citep{Boss:1997,Boss:1998,Durisen:2007,Zhu:2012}.

However, the exact value of the critical $\beta$ for fragmentation remains an open question. Non-convergence of the fragmentation criterion may arise from numerical artefacts \citep{Meru:2010,Lodato:2011,Meru:2012}, inherent stochasticity of fragmentation \citep[][]{Paardekooper:2012,Hopkins:2013}, the dimension (i.e. 2D vs. 3D) \citep{Young:2015} or the fact that there is no physical temperature floor in the $\beta$ cooling prescription \citep{Lin:2016}. Irradiation can be a main heating source in cool protoplanetary discs subject to GI, and thus may affect the fragmentation criterion \citep{Rice:2011}. It has also been suggested that a fragmentation criterion in terms of the $\alpha$ parameter \citep{Shakura:1973} may be more general than the cooling time $\beta$ \citep{Rice:2005}.

On the other hand, some authors have claimed that the cooling time $\beta$ is not necessarily the primary factor for fragmentation. \citet{Rogers:2012} proposed that the Hill radius plays an essential role in fragmentation; that is, fragmentation occurs when the width of a spiral density wave is less than the Hill radius, although the width itself may be determined by the balance between cooling and heating. \citet{Tsukamoto:2015} also found that fragmentation discs have narrower spiral density waves than non-fragmentation discs, and emphasised that the local minimum of Toomre's parameter inside the spiral density waves, $Q_\text{min}$, determines whether they fragment (for $Q_\text{min} \lesssim 0.2$) or not. \citet{Takahashi:2016}, based on a linear analysis, related the critical Toomre's parameter below which fragmentation occurs and the width of a density wave, and derived a fragmentation condition as $Q_\text{min} \lesssim 0.6$ for typical density waves in their global simulations.

In a series of papers \citep[][hereafter Paper I, and this paper]{Hirose:2017}, we have examined the fragmentation condition, as well as the gravito-turbulence, in an irradiated protoplanetary disc in the framework of a local shearing box. Given that there are many studies in the literature, our stance is as follows. Because temperature is one of the most important quantities that control GI (see eq. \ref{eq:toomre_condition}), correct thermodynamic analysis is essential to study the nonlinear evolution of GI in realistic protoplanetary discs. Therefore, we perform 3D radiation hydrodynamics simulations with a realistic opacity and a realistic equation of state (EOS), and include the irradiation heating by the central star. This is an extended work from \citet{Shi:2014}, who performed 3D local shearing box simulations using the $\beta$ cooling and simple optically-thin cooling prescriptions. The local shearing box has two physical parameters, the distance from the central star $r$ and the surface density $\Sigma$. In Paper I, we mainly studied the dependence on $\Sigma$ of the nonlinear outcome of GI at a single radius of $r = 50$ AU. It is therefore the goal of this paper to present the nonlinear outcome of GI in a relatively complete $\Sigma$-$r$ parameter space. Especially, we map out the regions in which the disc is laminar, turbulent, or fragmenting in the $\Sigma$-$r$ parameter space, and provide physical interpretations to such a phase diagram. In this sense, this work is also an extension of \citet{Johnson:2003}, who presented similar mapping based on 2D shearing box simulations but lacked a realistic EOS and did not consider irradiation heating. 

This paper is organised as follows. After we briefly describe our numerical methods in Section \ref{sec:methods}, we present the nonlinear outcome of GI and discuss the properties of gravito-turbulence as well as the fragmentation condition in Section \ref{sec:results}. In Section 4, we compare our results with previous studies and discuss some implications. Finally, we provide a summary in Section \ref{sec:summary}.

\begin{figure*}
  \includegraphics[width=0.72\textwidth]{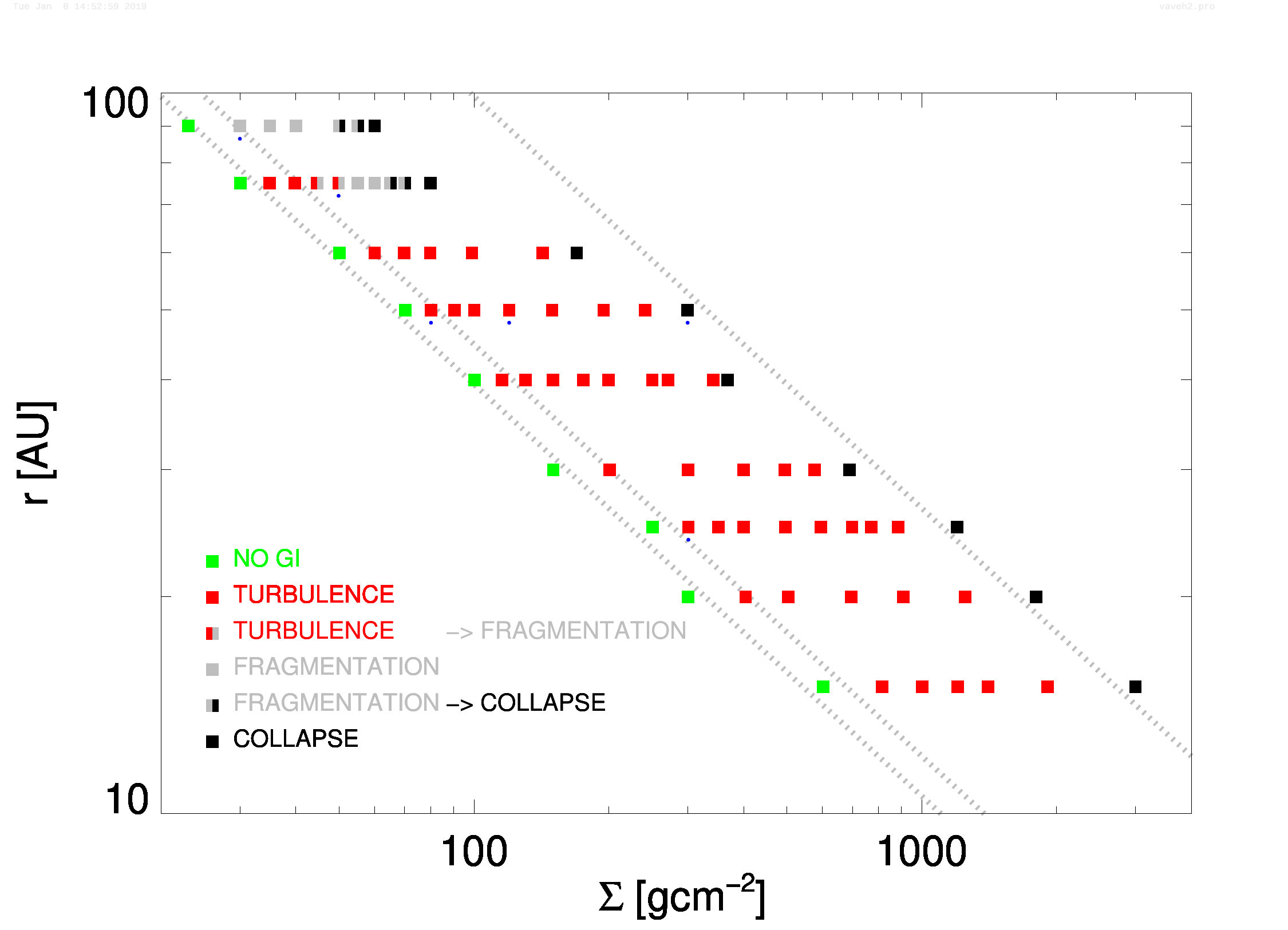}
  \caption{Nonlinear outcome of GI in the $\Sigma$-$r$ space.
    Green, red, red--grey, grey, grey--black, and black squares denote, respectively, no GI, turbulence, turbulence followed by fragmentation, fragmentation, fragmentation followed by runaway collapse, and runaway collapse.
    The dotted lines denote, respectively, $Q_0 = 1$, $0.8$, and $0.2$, from left to right.
    The squares with a dot below correspond to the runs shown in Figs. \ref{fig:self6004hh}, \ref{fig:self6200hh}, \ref{fig:self6008a}, \ref{fig:self7001hh}, \ref{fig:self1003ii}, and \ref{fig:self9000ii}, respectively.}
  \label{fig:phase_diagram}
\end{figure*}
\section{Methods}\label{sec:methods}

In this section, we explain the methods we used only briefly, because they are the same as those used in Paper I, which the reader may refer to for additional details. 

\subsection{Basic equations and numerical schemes}

The basic equations solved in our simulations are hydrodynamics equations with Poisson's equation for self-gravity and frequency-integrated angular-moment equations of the radiative transfer:
\begin{align}
  &\frac{\partial\rho}{\partial t} + \nabla\cdot(\rho\bm{v}) = 0, \\
  &\frac{\partial(\rho\bm{v})}{\partial t} + \nabla\cdot(\rho\bm{v}\bm{v}) =
  -\nabla p -\rho\nabla\Phi + \frac{{\kappa}_\text{R}\rho}{c}\bm{F}, \label{eq:motion}\\
  &\frac{\partial e}{\partial t} + \nabla\cdot(e\bm{v}) =
  -(\nabla\cdot\bm{v})p - \left(4\pi B(T) - cE\right){\kappa}_\text{P}\rho, \label{eq:energy_gas}\\
  &\frac{\partial E}{\partial t} + \nabla\cdot(E\bm{v}) =
  -\nabla\bm{v}:\mathsf{P} + \left(4\pi B(T) - cE\right){\kappa}_\text{P}\rho - \nabla\cdot\bm{F}, \label{eq:energy_rad}\\
  &\nabla^2\Phi = 4\pi G\rho, \label{eq:poisson}
\end{align}
where $\rho$ is the gas density, $e$ is the gas internal energy, $p$ is the gas pressure, $T$ is the gas temperature (assumed to be the same as the dust temperature), $E$ is the radiation
energy density, $\mathsf{P}$ is the radiation pressure tensor, $\bm{F}$ is the
radiation energy flux, $\bm{v}$ is the velocity field vector, $B(T) = \sigma_\text{B}T^4/\pi$ is the Planck
function ($\sigma_\text{B}$ is the Stefan-Boltzmann constant), and $c$ is the speed
of light.
The flux limited diffusion approximation was employed to close the angular-moment equations, where the first and second moments, $\bm{F}$ and $\mathsf{P}$, are related to the zeroth moment, $E$ \citep{Turner:2001}.

The EOS, $p = p(e,\rho)$ and $T = T(e,\rho)$, is an updated version of that used in \citet{Tomida:2012fb} in their star formation simulations. The Rosseland- and the Planck- mean opacity, $\kappa_\text{R}(\rho,T)$ and $\kappa_\text{P}(\rho,T)$, are the same as those used in \citet{Hirose:2015}, where the dust and gas opacity are taken from, respectively, \citet{Semenov:2003} and \citet{Ferguson:2005}. 

We used the shearing box approximation to model a local patch of an accretion disc 
as a co-rotating Cartesian frame $(x,y,z)$ with the linearized Keplerian shear flow, $\bm{v}_\text{K} \equiv -(3/2)\Omega x\hat{\bm{y}}$. 
The inertial forces in the co-rotating frame and the vertical component of the external gravity by the central star are 
added as source terms in the equation of motion (\ref{eq:motion}). 
Shearing-periodic, periodic, and outflow boundary conditions are applied to the boundaries in the $x$, $y$, and $z$ direction, respectively \citep{Hirose:2006}.

We employed ZEUS \citep{Stone92a} to solve the above equation set. An orbital 
advection algorithm \citep{Stone:2010} was implemented for accurate calculation in a wide shearing box.
Poisson's equation with the vacuum boundary condition in the $z$ direction was solved by Fast Fourier Transforms \citep{Koyama:2009}. The irradiation heating rate, evaluated by solving a time-independent radiative transfer equation (ignoring scattering), was added as a source term in equation (\ref{eq:energy_gas}). The nonlinear radiative transfer terms in the energy equations (\ref{eq:energy_gas}) and (\ref{eq:energy_rad}) were coupled to be solved time-implicitly using the Newton-Raphson method. The kinetic energy dissipating either numerically or physically was captured in the form of gas internal energy, which guaranteed conservation of the sum of the kinetic and internal energies \citep{Hirose:2006}.

\subsection{Parameters and the initial conditions}\label{sec:parameters}
A stratified shearing box has two physical parameters. One is the orbital frequency $\Omega = \sqrt{GM_*/r^3}$ [s$^{-1}$], which appears in the inertial force terms and the shearing periodic boundary condition. Here $M_*$ is the mass of the central star, $r$ is the distance from the central star, and $G$ is the gravitational constant. The other is the (horizontally-averaged) surface density $\Sigma$ [g cm$^{-2}$], which represents the amount of gas in the box. In our simulations, the value of $\Sigma$ varied from the initial value $\Sigma_0$ due to the outflow boundary condition as well as the density floor \citep[see][for details]{Hirose:2006}. However, because the relative difference was typically small (a few percent in one hundred orbits at largest), we do not explicitly distinguish $\Sigma$ and $\Sigma_0$ in this paper.

The parameters of irradiation heating are the energy flux $F_\text{irr} = (R_*/r)^2\sigma_\text{B}T_*^4$ [erg cm$^{-2}$ s$^{-1}$] and the grazing angle $\theta$, where $R_*$ and $T_*$ are, respectively, the radius and the effective temperature of the central star. We assumed that $T_* = 4000$ K, $M_* = 1M_\odot$, and $R_* = 1R_\odot$. Also, we fixed the grazing angle as $\theta = 0.02$ for simplicity because the main effect of the irradiation heating (i.e. setting a physical temperature floor near the midplane) only weakly depends on $\theta$ (see eq. (\ref{eq:T_0}) below).

The initial disc was set up to be isothermal and in hydrostatic equilibrium ignoring self-gravity, where a mean molecular weight $\mu = 2.38$ and adiabatic exponent $\gamma = 5/3$ were used. The isothermal temperature was evaluated using the radiative equilibrium disc model \citep[Equation~12a in ][]{Chiang97} as
\begin{align}
  T_0 = \left(\frac{\theta}{4}\right)^\frac14\left(\frac{R_*}{r}\right)^\frac12T_*.\label{eq:T_0}
\end{align}
The initial radiation field $E_0$ was assumed to be in thermal equilibrium with the gas, where $E_0 = (4\sigma_\text{B}/c)T_0^4$. The initial velocity field was the linearized Keplerian shear flow, whose $x$ and $z$ components were perturbed randomly up to 0.5\% of the local sound speed $c_\text{s} \equiv \sqrt{\Gamma \left(p/\rho\right)}$, where $\Gamma\equiv d\ln p/d\ln\rho$ is the generalised adiabatic exponent.

In all runs, the box size and the number of cells were set as $(L_x,L_y,L_z) = (24H,24H,12H)$ and $(N_x,N_y,N_z)=(128,128,64)$, respectively. Here and hereafter, the scale height of the initial isothermal disc $H \equiv \sqrt{2RT_0/(\mu\Omega^2)}$ is used as the unit length, where $R$ is the gas constant.

\begin{table}
 \caption{List of the selected five runs compared to see dependence of the nonlinear outcome on $\Sigma$ (runs S0, S1, and S2) and $r$ (runs R0, R1, and R2). Run S0 is identical to Run R1.}
 \label{table:run_list}
 \begin{tabular}{llllrl}
  \hline
  Label & Surface density $\Sigma$ & Radius $r$ & outcome & \multicolumn{2}{c}{Fig.\#} \\
  \hline
  S0 & $\Sigma_{0.82}$ ($80$ gcm$^{-2}$) & $50$ AU &turbulence& \ref{fig:self6004hh} &\multirow{3}{*}{\ref{fig:slicew_hikaku_sigma}}\\
  S1 & $\Sigma_{0.55}$ ($120$ gcm$^{-2}$)& $50$ AU &turbulence& \ref{fig:self6200hh} &\\
  S2 & $\Sigma_{0.22}$ ($300$ gcm$^{-2}$)& $50$ AU &runaway collapse& \ref{fig:self6008a}  &\\
  \hline
  R0 & $\Sigma_{0.74}$ ($300$ gcm$^{-2}$)& $25$ AU &turbulence& \ref{fig:self7001hh} &\multirow{3}{*}{\ref{fig:slicew_hikaku_radius}}\\
  R1 & $\Sigma_{0.82}$ ($80$ gcm$^{-2}$) & $50$ AU &turbulence& \ref{fig:self6004hh} &\\
  R2 & $\Sigma_{0.78}$ ($30$ gcm$^{-2}$) & $90$ AU &fragmentation& \ref{fig:self1003ii} &\\
  \hline
 \end{tabular}
\end{table}

\section{Results}\label{sec:results}

\begin{figure}
  \includegraphics[width=\columnwidth]{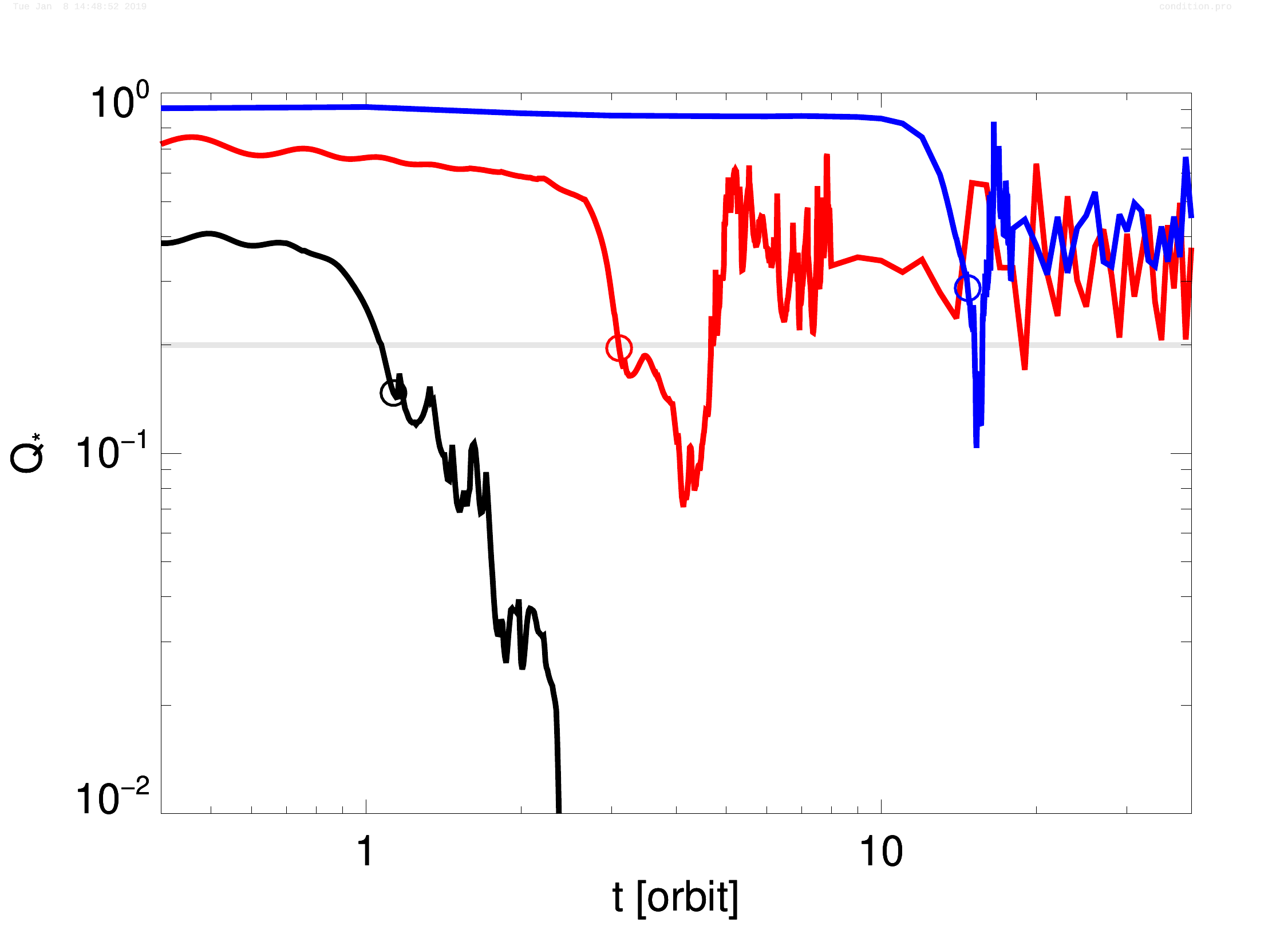}  
  \includegraphics[width=\columnwidth]{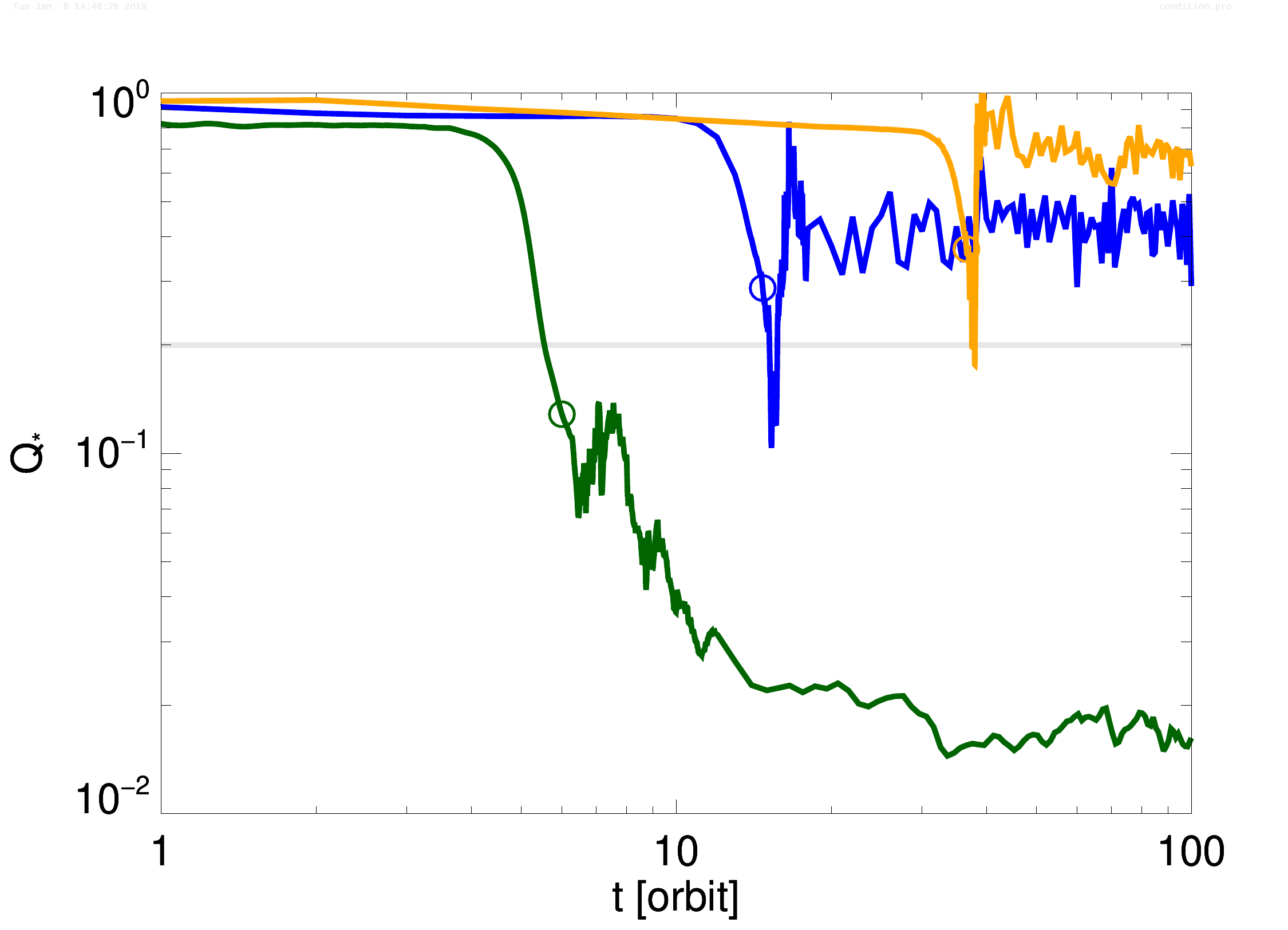} 
  \caption{Time evolution of the local Toomre's parameter $Q_\sm$. The upper panel compares runs S0 (blue), S1 (red) and S2 (black), whilst the lower panel compares runs R0 (orange), R1 (blue), and R2 (green); refer to Table \ref{table:run_list} for the labels of the runs. The open circle on each curve denotes the end of the growth of the axisymmetric self-gravitating density waves. The horizontal grey line denotes the value of $0.2$.}
  \label{fig:condition_hikaku_q}
\end{figure}
\begin{figure}
  \includegraphics[width=\columnwidth]{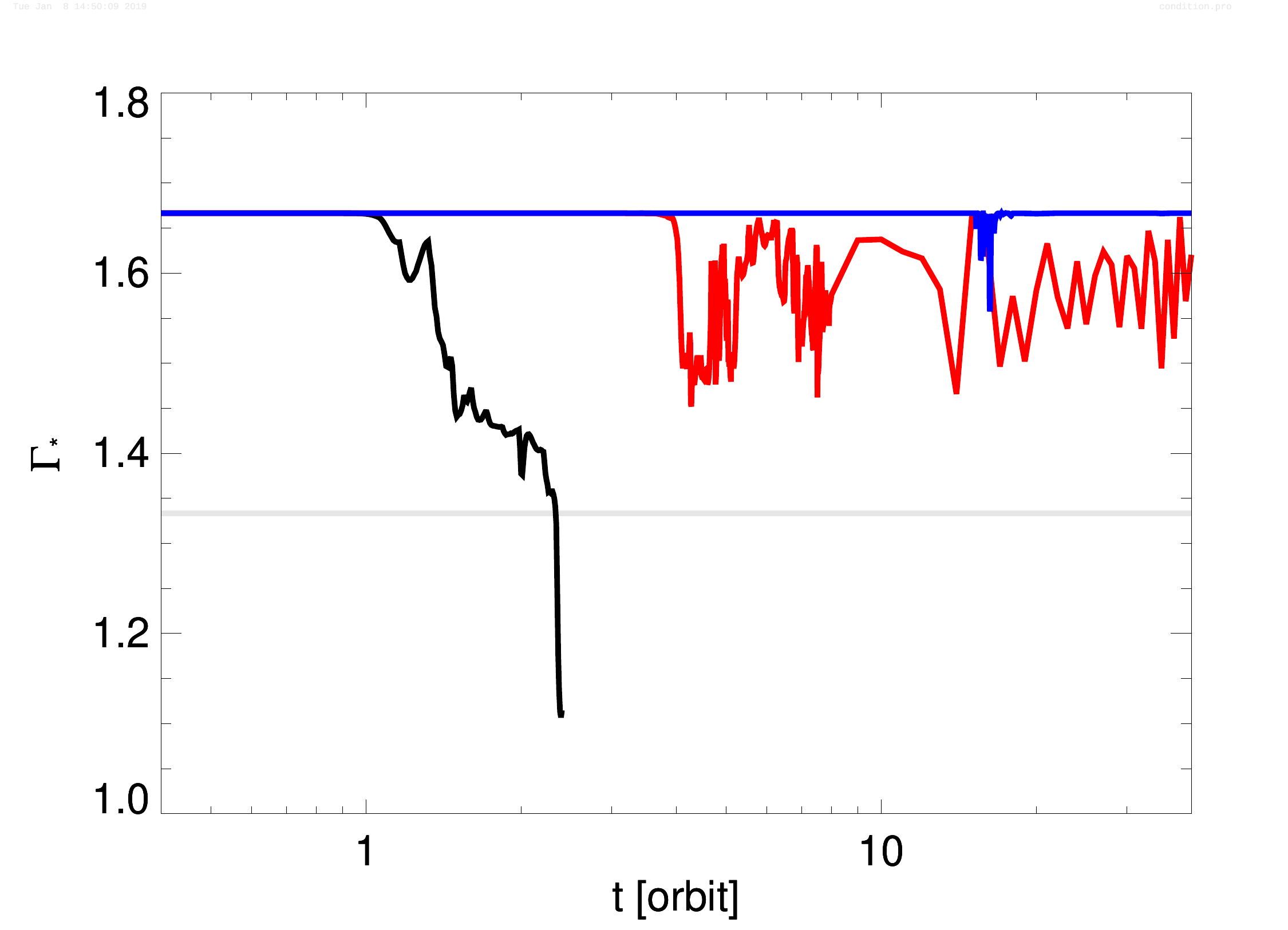} 
  \includegraphics[width=\columnwidth]{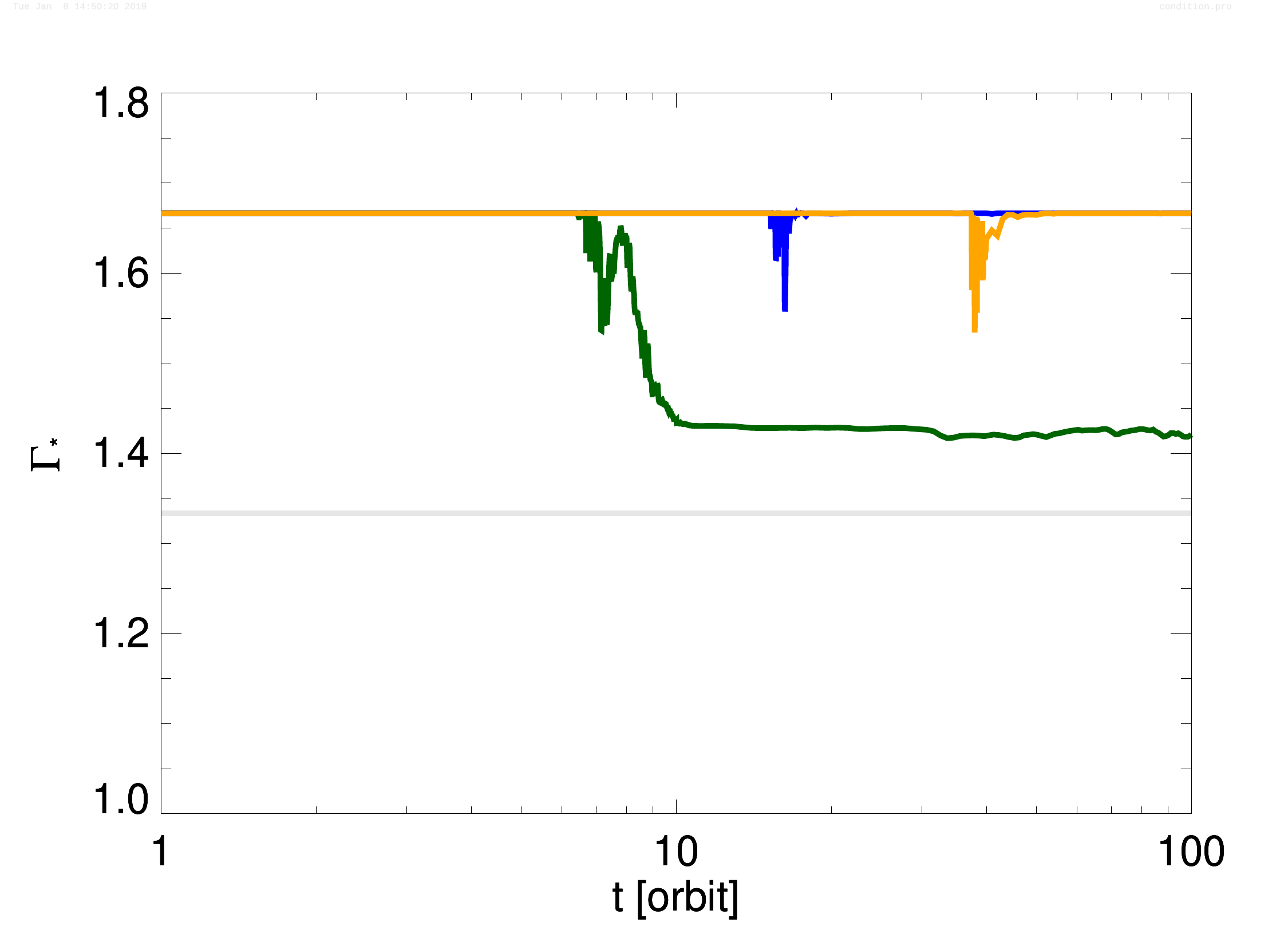} 
  \caption{Same as Fig. \ref{fig:condition_hikaku_q}, except for the local adiabatic exponent $\Gamma_\sm$. The horizontal grey line denotes the critical value of $\Gamma = 4/3$.}
  \label{fig:condition_hikaku_gamma}
\end{figure}

\subsection{Diagnostics}
For diagnostics, we use simple and density-weighted volume averages for a quantity $f(x,y,z)$, defined as
\begin{align}
&\left\llangle f\right\rrangle \equiv \dfrac{\int\left\langle f\right\rangle dz}{\int dz} &&\text{simple volume average},\\
&\left\llangle f\right\rrangle_\text{mid} \equiv \dfrac{\int\left\langle f\right\rangle\left\langle \rho\right\rangle dz}{\int\left\langle\rho\right\rangle dz} &&\text{density-weighted volume average},
\end{align}
where $\left\langle f\right\rangle \equiv {\iint f(x,y,z)dxdy}/{\iint dxdy}$ is the horizontal average.

Also, we define locally the Toomre's parameter and the normalised cooling time in the midplane, respectively, as
\begin{align}
  &Q(x,y) \equiv \dfrac{{c_\text{s}}(x,y)\kappa}{\pi G\sigma(x,y)},\\
  &\beta(x,y) \equiv \dfrac{e(x,y,z=0)\Omega}{-q^-(x,y,z=0)},\label{eq:beta_mid}
\end{align}
where $\sigma(x,y) \equiv \int\rho(x,y,z) dz$ is the local surface density, ${c_\text{s}}(x,y) \equiv {\int c_\text{s}(x,y,z)\rho(x,y,z)dz}/{\sigma(x,y)}$ is the density-weighted average of the sound speed, and $q^- \equiv -\kappa_\text{R}\rho(4\pi B(T) - cE)$ is the radiative cooling term in equation (\ref{eq:energy_gas}). In this paper, when evaluating Toomre's parameter $Q$, we assume $\kappa = \Omega$ (the Keplerian rotation) except in Section \ref{sec:fragmentation_conditions}.

As we are interested in the nonlinear outcome of GI, we often examine quantities evaluated at the cell where the self-gravitational energy, $E_\text{sg} \equiv \rho\phi/2$, takes the minimum value on the midplane. Hereafter, the subscript ``$_\sm$'' denotes a quantity at the cell of minimum $E_\text{sg}$ on the midplane; for example, $(x_\sm,y_\sm)$ denotes the horizontal position of that cell.

\subsection{Nonlinear development and outcome of GI}\label{sec:nonlinear_development}

We have run {\numberOfSimulations} simulations in total to explore the parameter ranges of $\Sigma_{1} \lesssim \Sigma \lesssim \Sigma_{0.2}$ and $15\text{ AU} \le r \le 90\text{ AU}$.\footnote{The specific value of $\Sigma$ as well as that of $r$ in each simulation are given in Appendix \ref{sec:averaging_period}.} Here, $\Sigma_{Q_0}$ denotes the surface density that corresponds to the initial Toomre's parameter $Q_0$; that is, $Q_0 = {c_\text{s}}_0\Omega/(\pi G\Sigma_{Q_0})$, where ${c_\text{s}}_0$ is the initial sound speed. The nonlinear outcome is summarised as a phase diagram in the $\Sigma$-$r$ space in Fig. \ref{fig:phase_diagram}.
In Paper I, we found at $r = 50$ AU that gravito-turbulence is sustained for a certain range of $\Sigma$ whilst GI is not driven below that range and runaway collapse occurs above it. Such dependence on $\Sigma$ can be seen at $r \lesssim 60$ AU. Specifically, GI is driven when $\Sigma$ exceeds $\sim\Sigma_1$ and runaway collapse occurs when $\Sigma$ exceeds $\sim\Sigma_{0.2}$. On the other hand, at $r = 90$ AU, when GI is driven, the outcome is always fragmentation (or runaway collapse) and no gravito-turbulence is sustained. The outcome at $r = 75$ AU is somewhat intermediate between $r \lesssim 60$ AU and $r = 90$ AU. 

Among the total {\numberOfSimulations} runs, we especially inspect in detail the five runs listed in Table \ref{table:run_list} to observe the dependence of the outcome on $\Sigma$ (runs S0, S1 and S2) as well as on $r$ (runs R0, R1 and R2). In Fig. \ref{fig:condition_hikaku_q}, we compare the time evolution of $Q_\sm$ amongst them, where $Q_\sm$ is the local Toomre's parameter evaluated at the cell of minimum $E_\text{sg}$ as
\begin{align}
  Q_\sm \equiv \frac{c_\text{s}(x_\sm,y_\sm)}{\pi G\sigma(x_\sm,y_\sm)}.\label{eq:local_Q}
\end{align}
The value of $Q_\sm$ in the final state is found to provide a good measure for distinguishing the outcome quantitatively as follows:
\begin{itemize}
  \item gravito-turbulence: $0.1 < Q_\sm$,
  \item fragmentation: $0.01 < Q_\sm < 0.1$,
  \item runaway collapse (fragmentation): $Q_\sm < 0.01$.
\end{itemize}
Here, runaway collapse is a special case of fragmentation in which gas pressure cannot stop the gravitational collapse due to softening of the EOS. Fig. \ref{fig:condition_hikaku_gamma} compares the time evolution of $\Gamma_\sm$ (the adiabatic exponent $\Gamma$ at the cell of minimum $E_\text{sg}$) amongst the five runs. In run S2, fragmentation was followed by runaway collapse because the core temperature exceeded the hydrogen dissociation temperature and thus $\Gamma$ dipped below the critical value of $4/3$. On the other hand, in run R2, although fragmentation did occur, runaway collapse did not occur and a pressure-supported clump survived because $\Gamma$ remained well above the critical value owing to insufficient rise of the core temperature. Once runaway collapse occurred, we stopped the calculation because the following evolution at smaller scales could not be treated in our simulation with a fixed grid.

\begin{figure*}
  \includegraphics[width=0.72\textwidth]{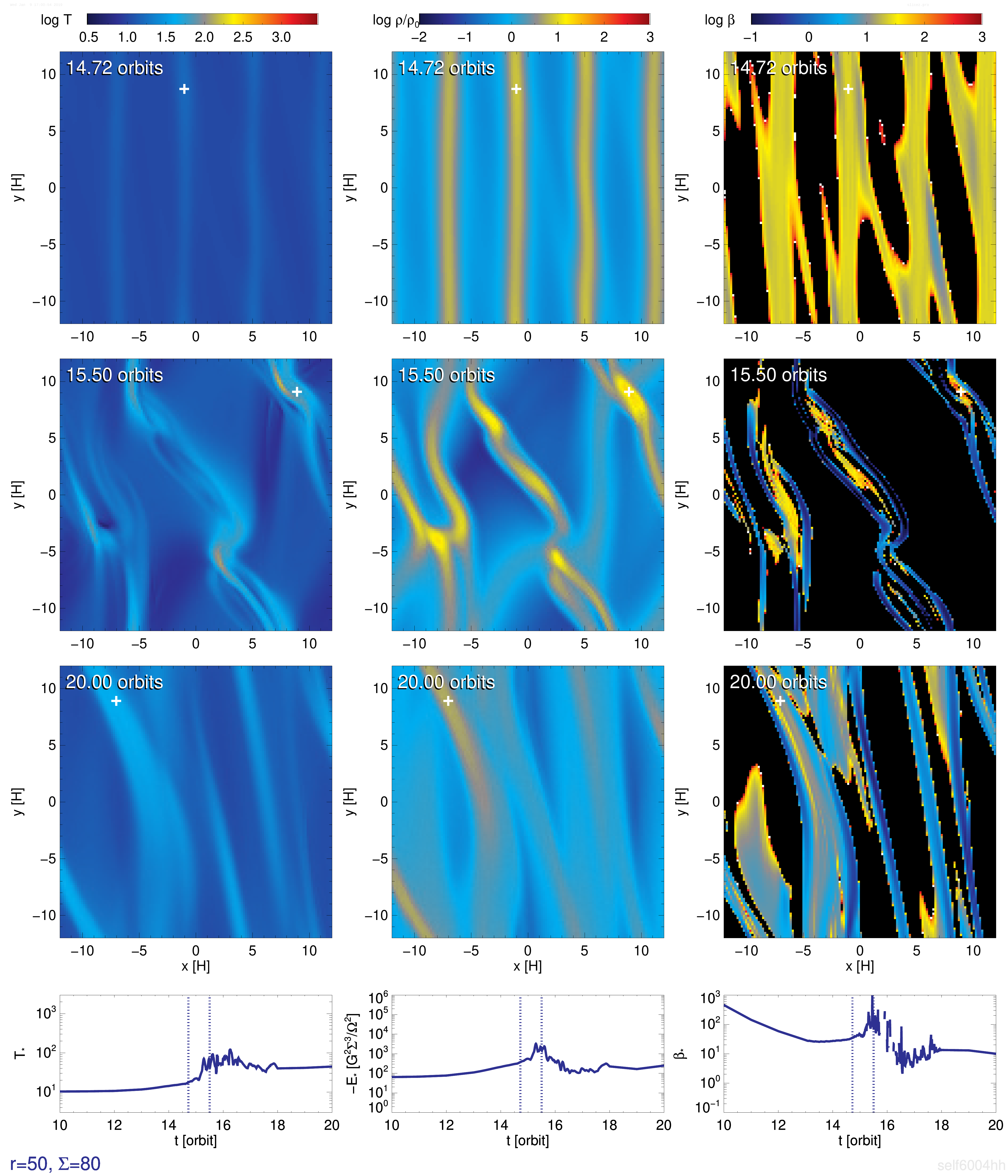} 
  \caption{Selected snapshots of $T(x,y,z=0)$ (left), $\rho(x,y,z=0)/\rho_0$ (middle), and $\beta(x,y)$ (right) in run S0 (identical to run R1), where $\rho_0$ is the initial midplane density. At the bottom, time evolutions of $T(x_\sm,y_\sm,z=0)$ (left), $-E_\text{sg}(x_\sm,y_\sm,z=0)$ (middle), and $\beta(x_\sm,y_\sm)$ (right) are shown respectively, where $(x_\sm,y_\sm)$ denotes the horizontal position of the cell of the minimum $E_\text{sg}$ on the midplane whilst the vertical dotted lines indicate the selected three instances respectively. In the logarithmic plot of $\beta_\sm$, negative values are not shown. The cell of the minimum $E_\text{sg}$ is shown as a white cross in the snapshots.}
  \label{fig:self6004hh}
\end{figure*}

\begin{figure*}
  \includegraphics[width=0.72\textwidth]{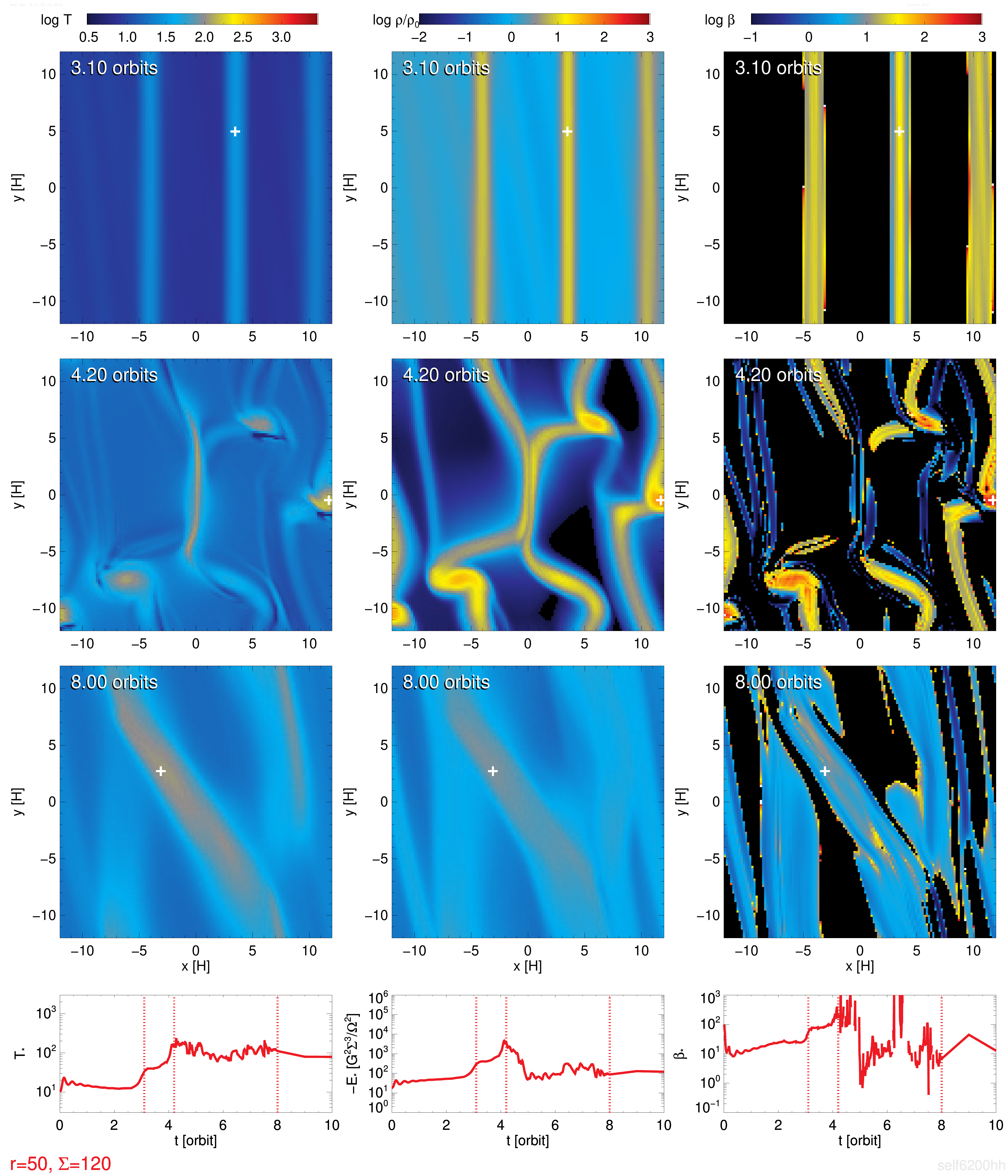} 
  \caption{Same as Fig. \ref{fig:self6004hh}, except for run S1}
  \label{fig:self6200hh}
\end{figure*}

\begin{figure*}
  \includegraphics[width=0.72\textwidth]{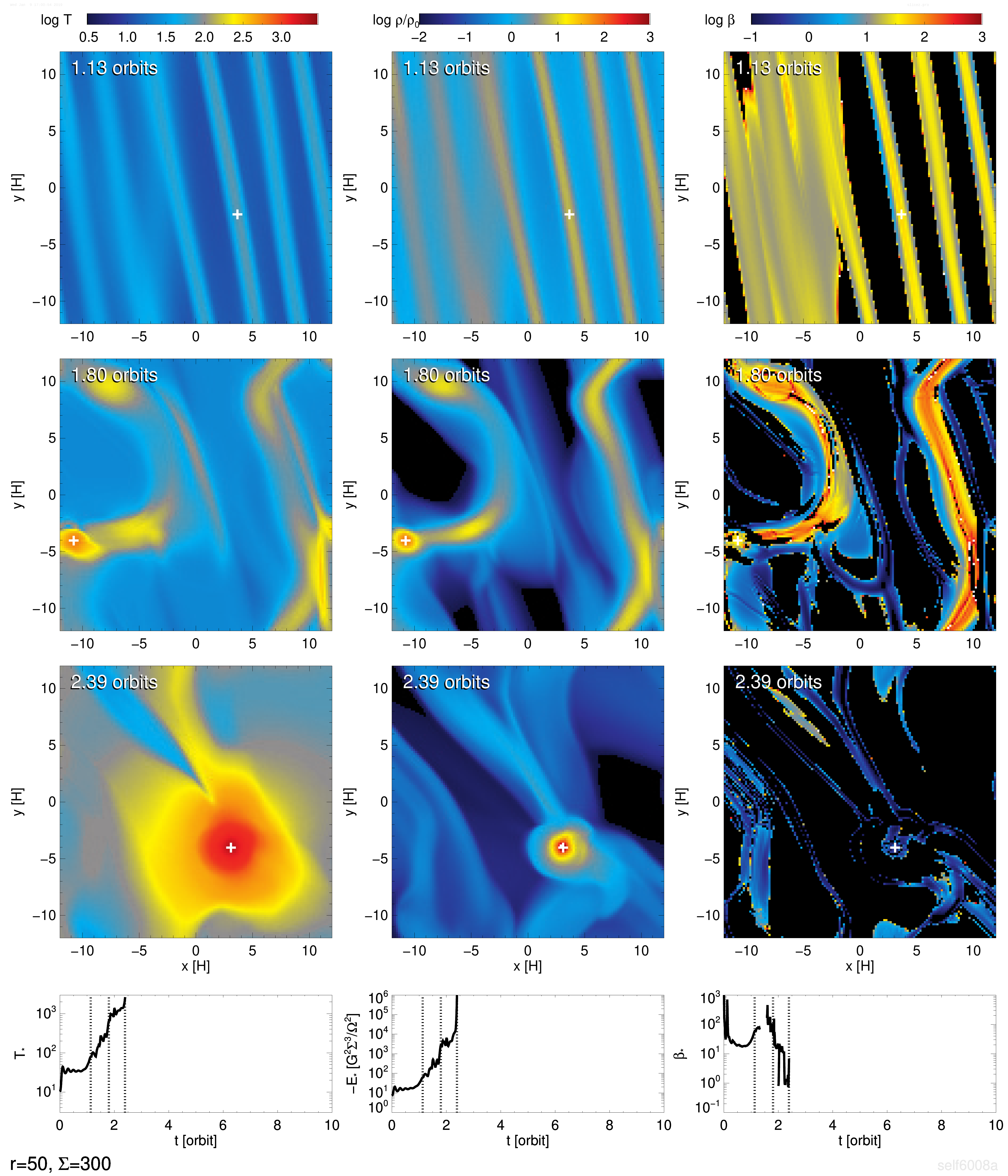} 
  \caption{Same as Fig. \ref{fig:self6004hh}, except for run S2}
  \label{fig:self6008a}
\end{figure*}

\begin{figure*}
  \includegraphics[width=0.72\textwidth]{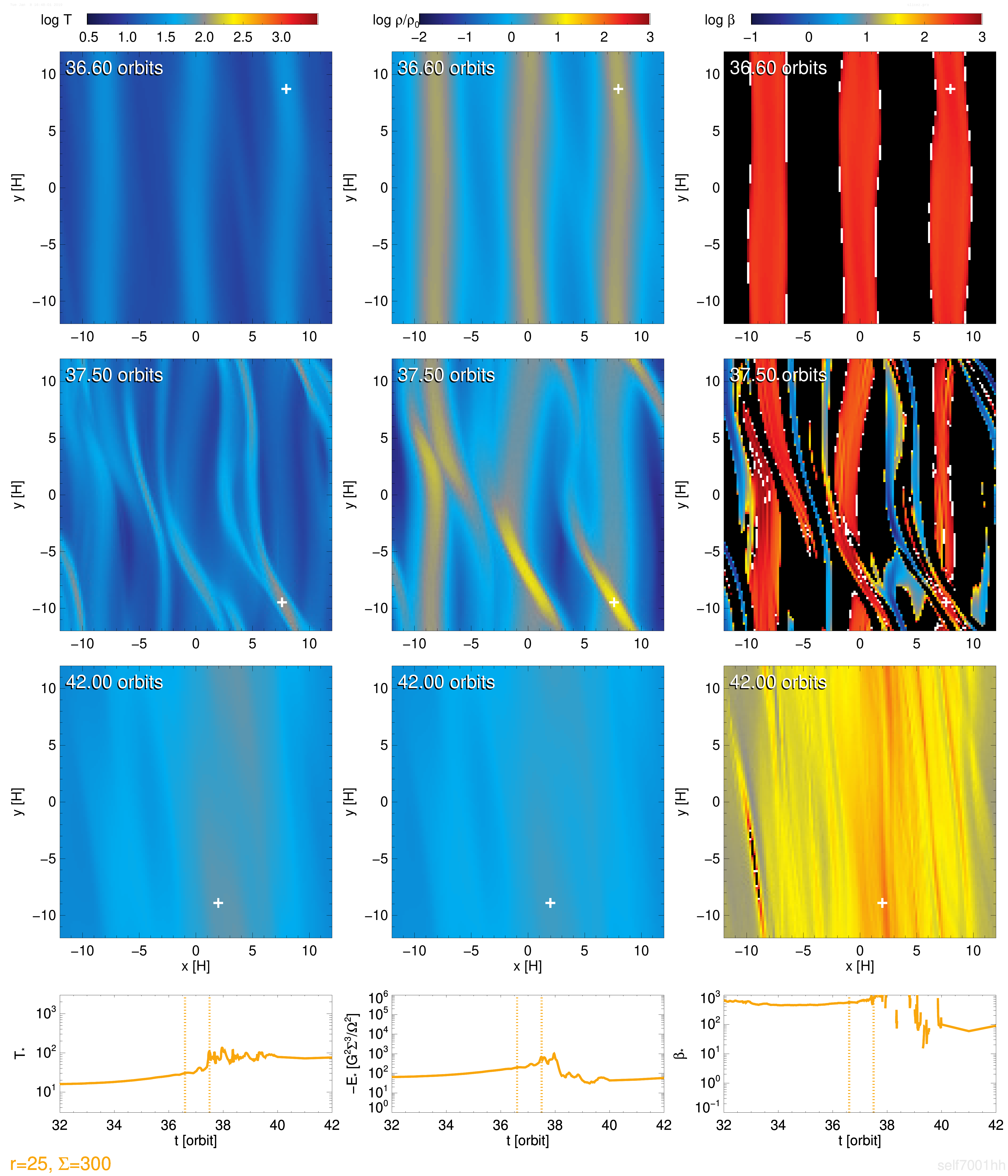} 
  \caption{Same as Fig. \ref{fig:self6004hh}, except for run R0}
  \label{fig:self7001hh}
\end{figure*}

\begin{figure*}
  \includegraphics[width=0.72\textwidth]{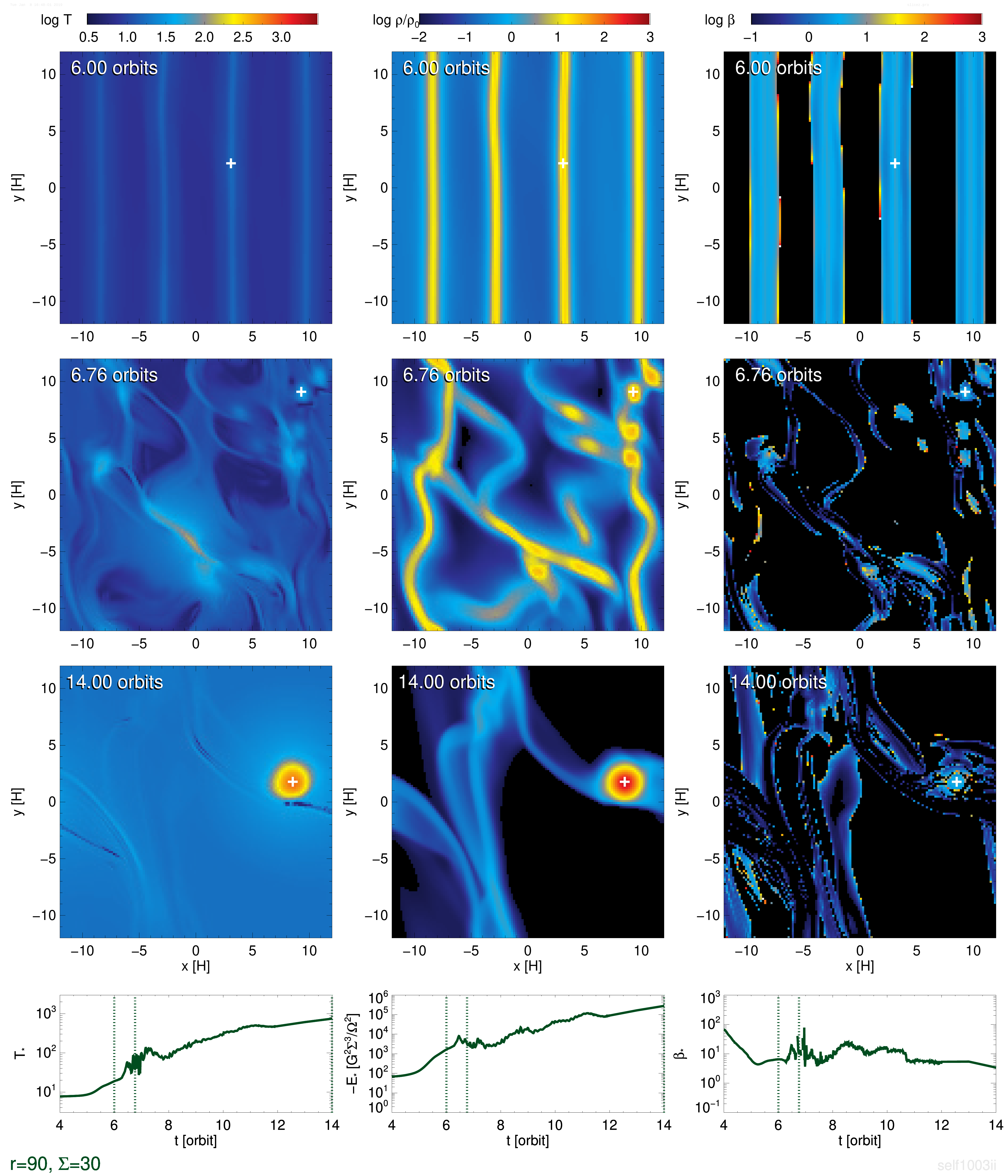} 
  \caption{Same as Fig. \ref{fig:self6004hh}, except for run R2}
  \label{fig:self1003ii}
\end{figure*}

Regardless of the outcome, the nonlinear evolution of GI followed the same steps:
\begin{enumerate}
\item destabilisation of the initial laminar flow, \label{step:1}
\item growth of almost axisymmetric density waves, \label{step:2}
\item non-axisymmetric destabilisation of the density waves, \label{step:3}
\item collapse of the density waves into a transient phase, \label{step:3.5}
\item the final outcome. \label{step:4}
\end{enumerate}
In Figs. \ref{fig:self6004hh}, \ref{fig:self6200hh}, \ref{fig:self6008a}, \ref{fig:self7001hh}, and \ref{fig:self1003ii}, we show time series snapshots of gas temperature $T(x,y,z=0)$, density $\rho(x,y,z=0)$, and the cooling time $\beta(x,y)$ for the selected five runs. In each figure, the top row corresponds to the epoch when the non-axisymmetric deformation of the almost axisymmetric density waves becomes clear by eye-measurement (step \ref{step:3}). The middle row shows the following transient phase where complicated interactions between density waves and clumps are apparent (step \ref{step:3.5}), and the bottom row shows the final outcome (step \ref{step:4}).

Here we note that the almost axisymmetric density waves in step \ref{step:2} are not simple nonlinear manifestations of the initial most unstable modes in step \ref{step:1}; rather, they emerged as a result of nonlinear interactions between the initial unstable modes. As the axisymmetric density waves grew, their Toomre's parameter decreased and they became strongly self-gravitating (Fig. \ref{fig:condition_hikaku_q}). Eventually, without exception, they became unstable against non-axisymmetric perturbations (step \ref{step:3}) and collapsed into the transient phase (step \ref{step:3.5}). We will discuss in detail the non-axisymmetric instability of the density waves in Section \ref{sec:fragmentation_conditions}.

Hereafter, we refer to the almost axisymmetric density waves in step \ref{step:2} simply as axisymmetric density waves, omitting ``almost'', for simplicity.


\subsection{Gravito-turbulence}

Gravito-turbulence is the state where turbulent dissipation and radiative cooling balance, which was established for a finite range of $\Sigma$ at $r \lesssim 75$ AU. In this section, we examine how the properties of the gravito-turbulence depend on $\Sigma$ and $r$. Quantities that will be discussed in this section are the ones time-averaged for a period in which the gravito-turbulence is sustained (see Appendix \ref{sec:averaging_period} for the period). The time interval when recording the numerical data was $0.01$ orbits, except for the bottom panel in Fig. \ref{fig:sigma_ave}, where the interval was $1$ orbit.

\subsubsection{Cooling time $\beta$ and Toomre's parameter $Q$}\label{sec:cooling_time}
\begin{figure}
  \includegraphics[width=\columnwidth]{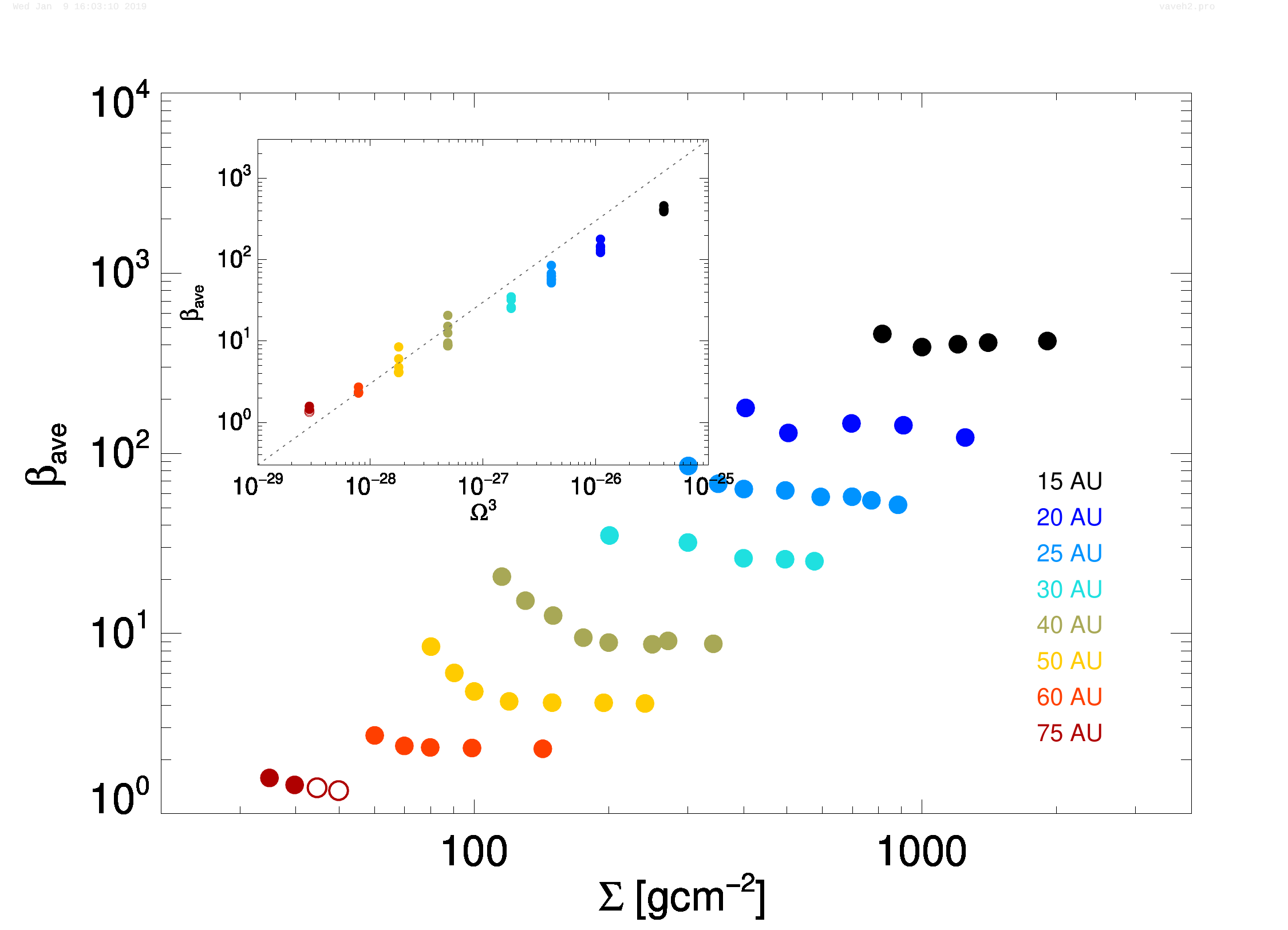} 
  \includegraphics[width=\columnwidth]{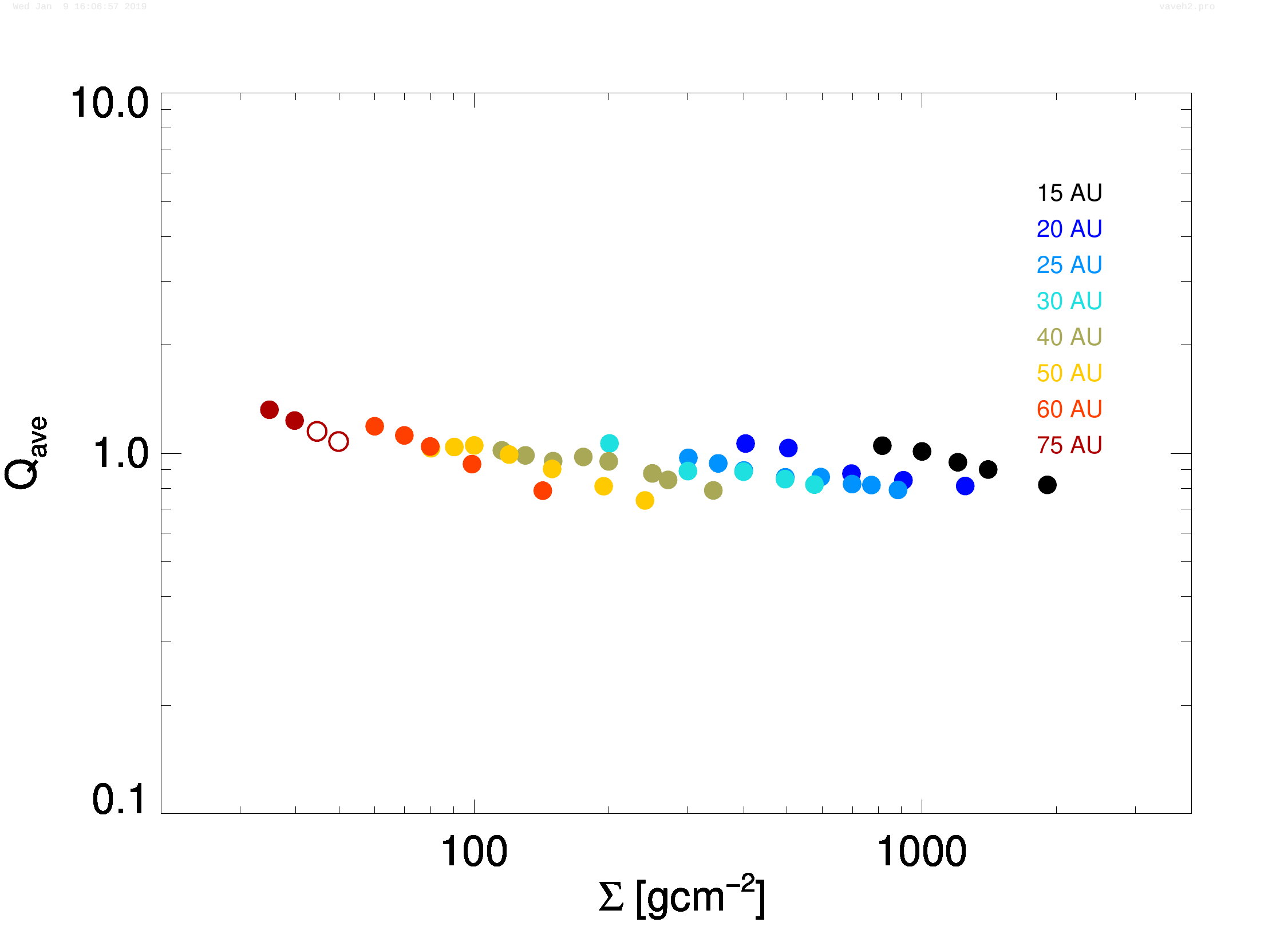} 
  \includegraphics[width=\columnwidth]{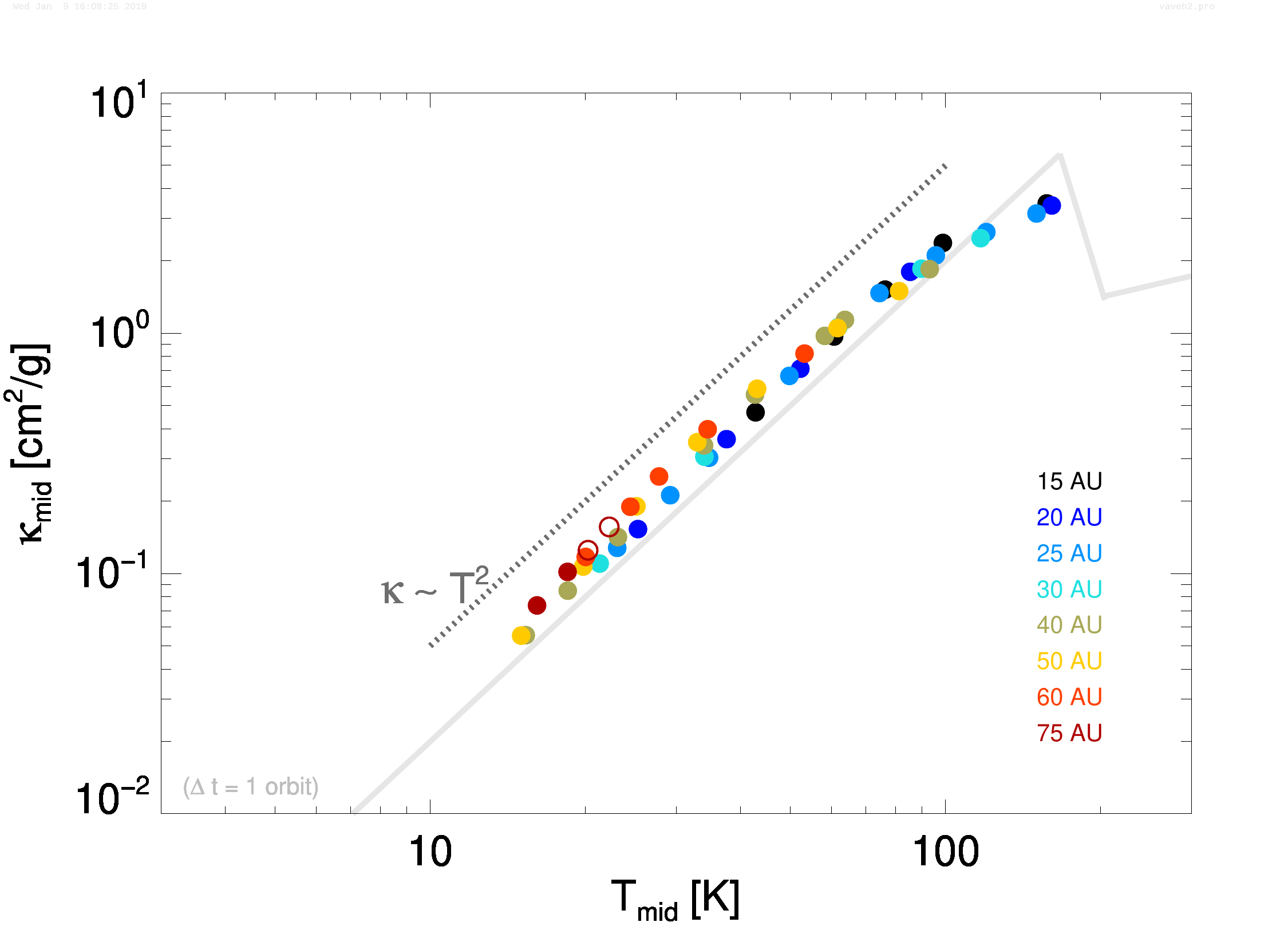} 
  \caption{Space-averaged cooling time $\beta_\text{ave}$ (top) and Toomre's parameter $Q_\text{ave}$ (middle) versus $\Sigma$, and Rosseland-mean opacity $\left\llangle\kappa\right\rrangle_\text{mid}$ versus gas temperature $\left\llangle T\right\rrangle_\text{mid}$ (bottom). The inset in the top panel shows $\beta_\text{ave}$ versus $\Omega$ (the dotted line denotes $\propto\Omega^3$). In the bottom panel, the dotted line denotes $\propto \left\llangle T\right\rrangle_\text{mid}^2$, whilst the grey curve denotes the opacity model used in \citet{Clarke:2009}. The different colours correspond to the different radii. The transition runs from gravito-turbulence to fragmentation are shown as open circles.}
  \label{fig:sigma_ave}
\end{figure}

In the top panel of Fig. \ref{fig:sigma_ave}, the space-averaged cooling time defined as
\begin{align}
  &\beta_\text{ave} \equiv \dfrac{\left\llangle e\right\rrangle_\text{mid}\Omega}{-\left\llangle q^-\right\rrangle_\text{mid}}
\end{align}
is shown in terms of $\Sigma$ and $r$. At each radius, $\beta_\text{ave}$ is almost constant except in some small $\Sigma$ cases, where it is relatively large because extra heating by irradiation raised the midplane thermal energy $\left\llangle e\right\rrangle _\text{mid}$. On the other hand, $\beta_\text{ave}$ strongly depends on $r$, as explicitly shown in the inset. Specifically, $\beta_\text{ave}$ is as small as $\sim 1.5$ at $r = 75$ AU, whilst it is as large as $\sim 400$ at $r = 15$ AU. As shown in the middle panel of Fig. \ref{fig:sigma_ave}, the space-averaged Toomre's parameter defined as
 \begin{align}
   &Q_\text{ave} \equiv \dfrac{\left\llangle {c_\text{s}}\right\rrangle _\text{mid}\Omega}{\pi G\Sigma}
 \end{align}
 is around unity for all runs, and it depends on $\Sigma$ and $r$ only weakly.
 
The strong dependence of $\beta_\text{ave}$ on $r$ is derived as follows \citep[e.g.][]{Clarke:2009,Paardekooper:2012}. Because the disc is optically thick, the cooling time is evaluated as the vertically integrated thermal energy $T_\text{mid}\Sigma$ divided by the radiative diffusion cooling rate $(T_\text{mid}^4/\kappa_\text{R}\Sigma)$, where $T_\text{mid}$ represents the midplane temperature. Then,
\begin{align}
&\beta \propto \frac{\left({T_\text{mid}}\Sigma\right)\Omega}{T_\text{mid}^4/\kappa_\text{R}\Sigma} \propto T_\text{mid}^{-1}\Sigma^2\Omega \propto \left(Q^{-2}\Omega^2\Sigma^{-2}\right)\Sigma^2\Omega \propto Q^{-2}\Omega^3\Sigma^0. \label{eq:beta_dependence}
\end{align}
Here we used the dependence of opacity on temperature $\kappa_\text{R} \propto T^2$ as shown in the bottom panel of Fig. \ref{fig:sigma_ave} as well as the definition of Toomre's parameter $Q \propto T_\text{mid}^{1/2}\Omega\Sigma^{-1}$, to eliminate $T_\text{mid}$. As we stated above, $Q_\text{ave}$ only weakly depends on $\Sigma$ and $r$. Therefore, if we ignore $Q$ in equation (\ref{eq:beta_dependence}), it becomes $\beta \propto \Omega^3\Sigma^0$, which is roughly consistent with the dependence shown in the inset of the top panel.

\subsubsection{Shear stress and $\alpha$}\label{sec:stress_and_alpha}
\begin{figure}
  \includegraphics[width=\columnwidth]{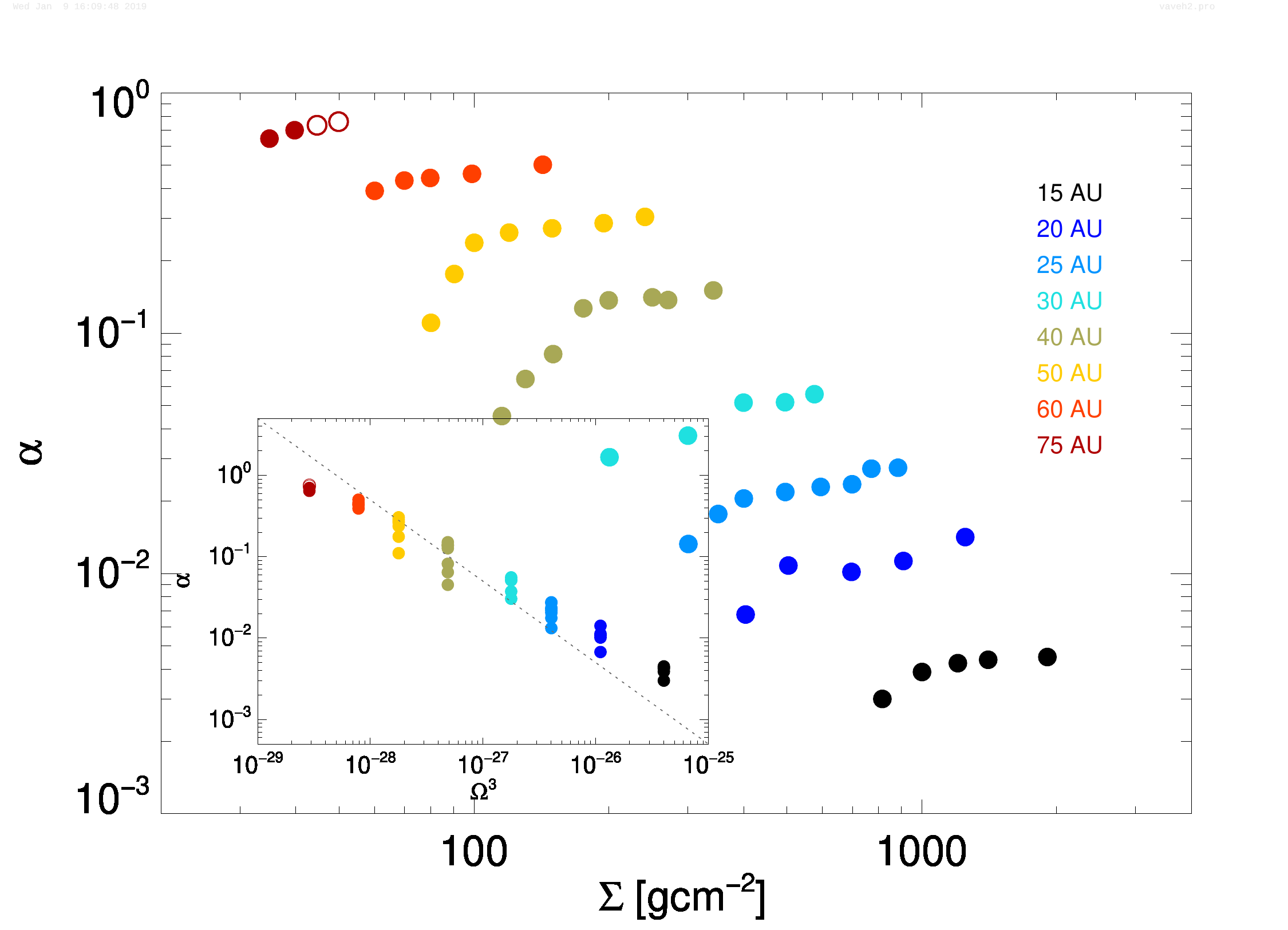} 
  \includegraphics[width=\columnwidth]{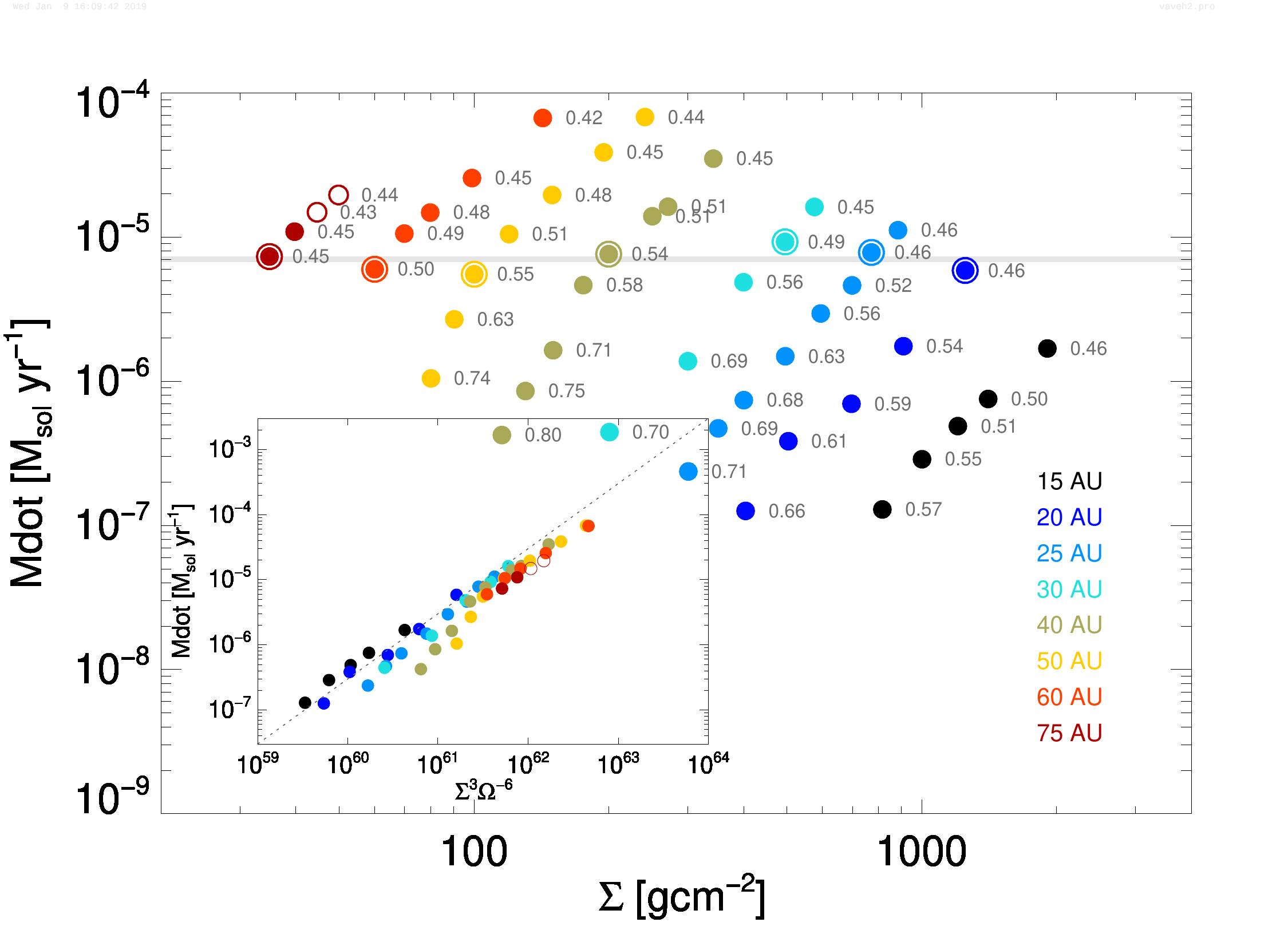} 
  \caption{The $\alpha$ parameter versus $\Sigma$ (upper) and the accretion rate $\dot{M}$ versus $\Sigma$ (lower). In the lower panel, the small number at the right of a symbol denotes the fraction of the gravitational stress to the total stress. The inset in the upper panel shows $\alpha$ versus $\Omega^3$ (the dotted line denotes $\propto \Omega^{-3}$), whilst that in the lower panel shows $\dot{M}$ versus $\Sigma^3\Omega^{-6}$ (the dotted line denotes $\propto \Sigma^3\Omega^{-6}$). The usage of colours and symbols is the same as in Fig. \ref{fig:sigma_ave}. In the bottom panel, the filled circles with a larger open circle are the solutions used in constructing a steady accretion disc model at $\dot{M} = 7\times10^{-6}$ $M_\odot$yr$^{-1}$ (denoted by the grey horizontal line) in Section \ref{sec:steady_accretion}.}
  \label{fig:sigma_mdot}
\end{figure}

The upper panel of Fig. \ref{fig:sigma_mdot} shows the dependence on $\Sigma$ and $r$ of the $\alpha$ parameter defined as
\begin{align}
  &\alpha \equiv \dfrac{\int\left\langle W_{xy}\right\rangle dz}{\int\left\langle P_\text{thermal}\right\rangle dz},
\end{align}
where the thermal pressure $P_\text{thermal}$ is the sum of the gas and radiation pressures, and the shear stress $W_{xy}$ is the sum of the gravitational and Reynolds shear stresses. The value of $\alpha$ widely ranges from $\sim 4\times10^{-3}$ at $r = 15$ AU to $\sim 0.7$ at $r = 75$ AU, and is roughly proportional to $\Omega^{-3}$, as shown in the inset. 
Comparing the upper panel of Fig. \ref{fig:sigma_mdot} with the top panel of Fig. \ref{fig:sigma_ave}, the dependence of $\alpha$ on $\Sigma$ and $r$ is almost opposite to that of $\beta_\text{ave}$. This is expected from the thermal balance condition, in that the vertically-integrated cooling rate $\int(\left\langle e\right\rangle + \left\langle E\right\rangle) dz\Omega/\beta_\text{ave}$ equals the vertically-integrated stress work $\frac32\Omega\int\left\langle W_{xy}\right\rangle dz$, which requires that $\alpha \propto \beta_\text{ave}^{-1}$ \citep{Gammie:2001}. 

Because the stress work $\frac32\Omega\int\left\langle W_{xy}\right\rangle dz$ is equated to the release rate of gravitational energy $\frac{3}{4\pi}\dot{M}\Omega^2$, the mass accretion rate $\dot{M}$ can be evaluated as
\begin{align}
  &\dot{M} = \dfrac{\int\left\langle W_{xy}\right\rangle dz}{\Omega/2\pi}\label{eq:mdot_stress},
\end{align}
which strongly depends on both $\Sigma$ and $r$, as shown in the lower panel of Fig. \ref{fig:sigma_mdot}.
The dependence of $\dot{M}$ on $\Sigma$ and $r$ can be derived as
\begin{align}
  \dot{M} \propto \alpha(T_\text{mid}\Sigma)\Omega^{-1} \propto \Omega^{-3}\left((Q^2\Omega^{-2}\Sigma^2)\Sigma\right)\Omega^{-1} \propto Q^2\Omega^{-6}\Sigma^3, \label{eq:mdot_scaling}
\end{align}
where $\alpha \propto \Omega^{-3}$ is substituted. Again, if we ignore the very weak dependence of Toomre's parameter on $\Sigma$ and $\Omega$, eq. (\ref{eq:mdot_scaling}) becomes $\dot{M} \propto \Sigma^3\Omega^{-6}$, which is roughly confirmed in the inset of the lower panel. Eq. (\ref{eq:mdot_scaling}) also states that the gravito-turbulence can sustain accretion flows of larger $\dot{M}$ at larger radii. Specifically, we note that the maximum accretion rate of $\dot{M} \sim 10^{-4}$ $M_\odot$yr$^{-1}$ is realised at $r = 50\sim 60$ AU, whilst the minimum accretion rate of $\dot{M} \sim 10^{-7}$ $M_\odot$yr$^{-1}$ is realised at $r = 15\sim 20$ AU. 

Also, the fraction of the gravitational stress to the total stress is generally around $\sim 0.5$ with a slight decrease with $\Sigma$ at each radius, as shown in the lower panel of Fig. \ref{fig:sigma_mdot}.

\subsubsection{Time variations}\label{sec:time_variation}
\begin{figure}
  \includegraphics[width=\columnwidth]{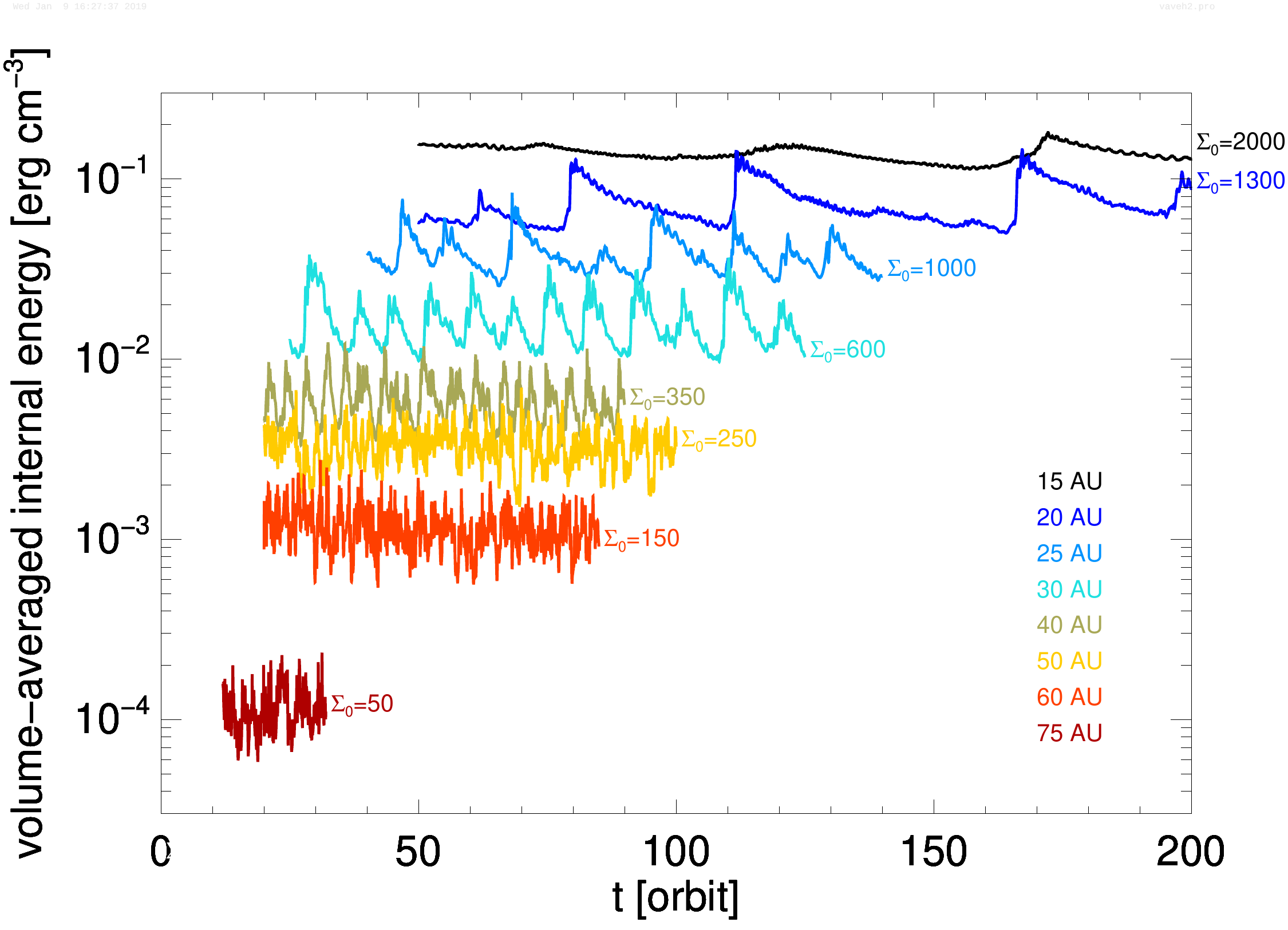} 
  \includegraphics[width=\columnwidth]{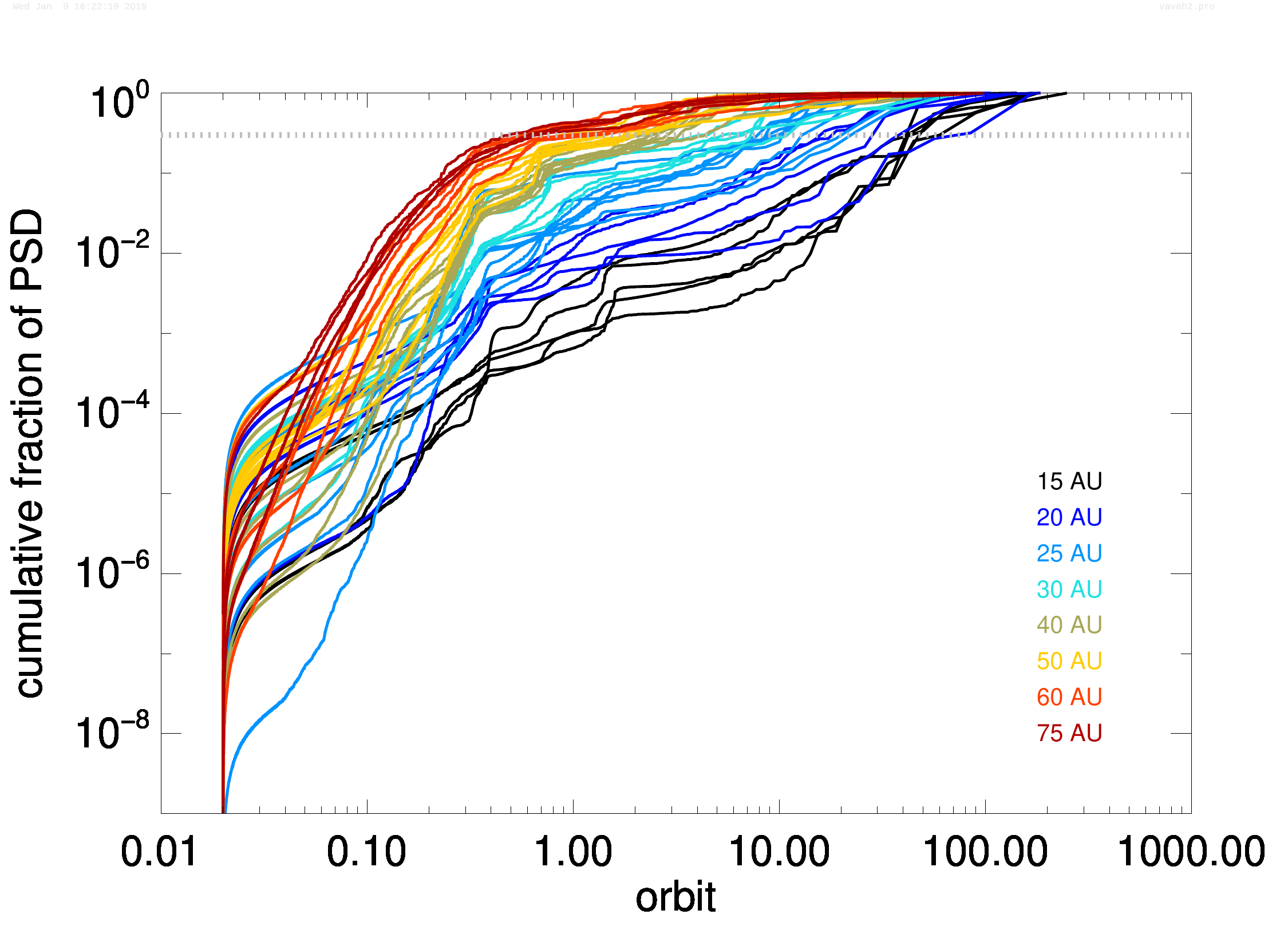} 
  \includegraphics[width=\columnwidth]{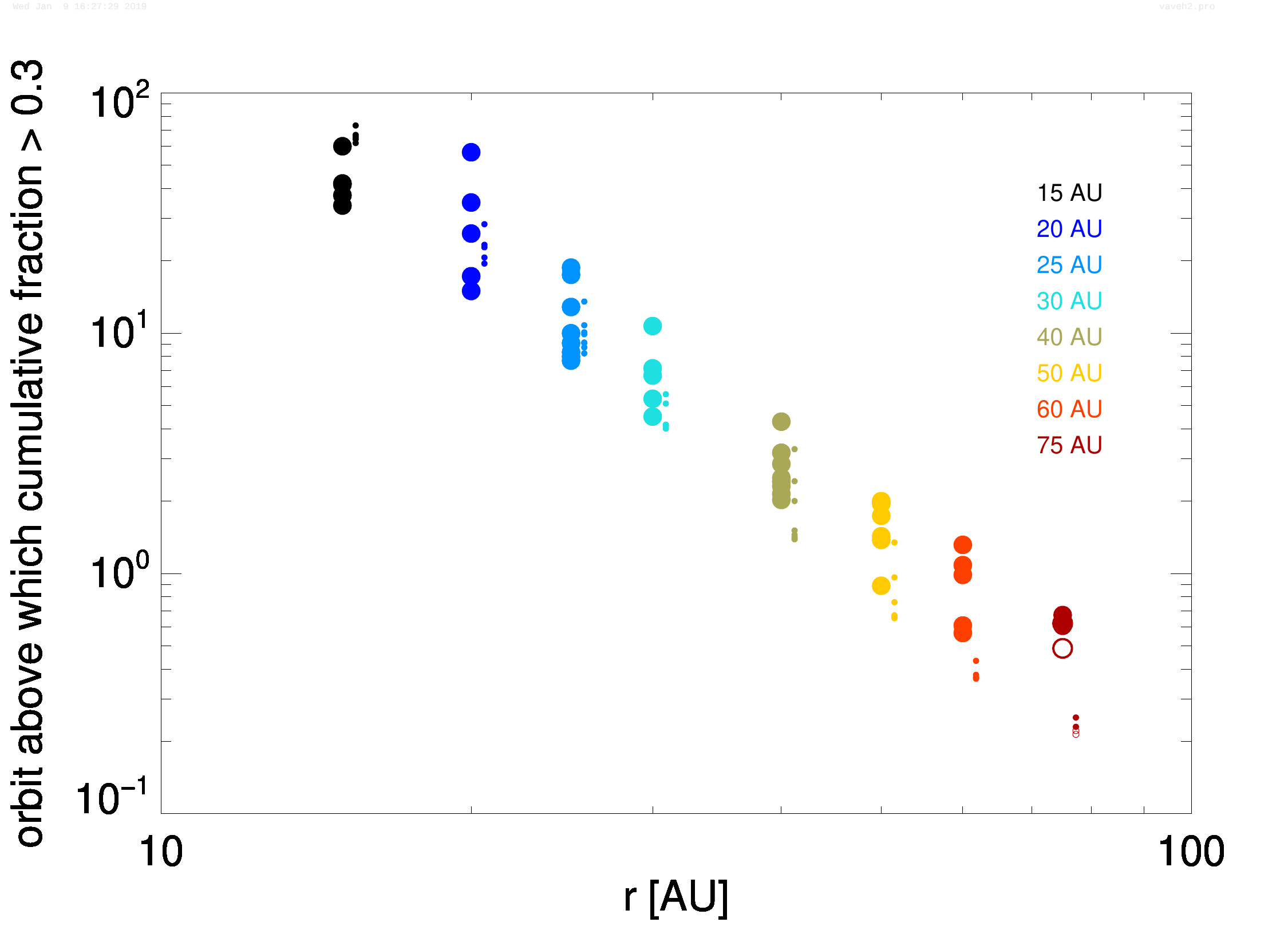} 
  \caption{Time variations of $\left\llangle e\right\rrangle$ of selected runs that sustain gravito-turbulence at different radii (top), cumulative fraction of the power spectral density (PSD) of $\left\llangle e\right\rrangle$ versus orbit for all runs that sustain gravito-turbulence (middle), and the orbit below which the cumulative fraction of PSD shown in the middle panel exceeds a critical value of $0.3$ versus radius (bottom). The dotted line in the middle panel denotes the critical value. In the bottom panel, the small dots at each radii show $\beta_\text{ave}$. The usage of colours and symbols is the same as in Fig. \ref{fig:sigma_ave}.}
  \label{fig:fftcum}
\end{figure}

In this section, we examine the temporal behaviour in the gravito-turbulence at different radii. The top panel of Fig. \ref{fig:fftcum} shows time variations of volume-averaged internal energy $\left\llangle e\right\rrangle$ of a representative run at different radii. At larger radii ($r \gtrsim 50$ AU), the time variation appears stochastic, but, as the radius decreases, it becomes more quasi-periodic, with longer periodicity. 

To quantify the typical time scale of variation of $\left\llangle e\right\rrangle$, we computed the one-sided power spectral density of $\left\llangle e\right\rrangle$,
\begin{align}
 & P(f) \equiv \left|\int_0^\infty \left\llangle e\right\rrangle e^{2\pi ift}dt\right|^2 + \left|\int_0^\infty \left\llangle e\right\rrangle e^{2\pi i(-f)t}dt\right|^2,
\end{align}
where $f$ is the frequency, and then take its cumulative fraction
\begin{align}
  & C(f) \equiv \dfrac{\int_f^\infty P(f')df'}{\int_0^\infty P(f')df'},
\end{align}
where $P(f=0)$ is excluded in the total power (the denominator). The result is shown in the middle panel of Fig. \ref{fig:fftcum}, and we observe a clear trend where, as the radius decreases, the power at longer periods provides a major contribution to the total power.

To quantify the trend above, we measure a frequency $f_0$ below which the cumulative fraction exceeds a critical value of $0.3$ (that is, $C(f) > 0.3$ for $f < f_0$), supposing that the inverse of the frequency $f_0$ represents the typical time scale of $\left\llangle e\right\rrangle$. The choice of the critical value is arbitrary, but here it is chosen so that the trend can be best observed. The bottom panel of Fig. \ref{fig:fftcum} shows that the typical time scale $1/f_0$ has a fairly strong dependence on $r$. Also, the figure shows that it has good correlation with the space-averaged cooling time $\beta_\text{ave}$, indicating that the time variation of $\left\llangle e\right\rrangle$ is mainly determined by the space-averaged cooling time.

\subsection{Boundaries between gravito-turbulence and fragmentation}\label{sec:fragmentation_condition}
\begin{figure}
  \includegraphics[width=\columnwidth]{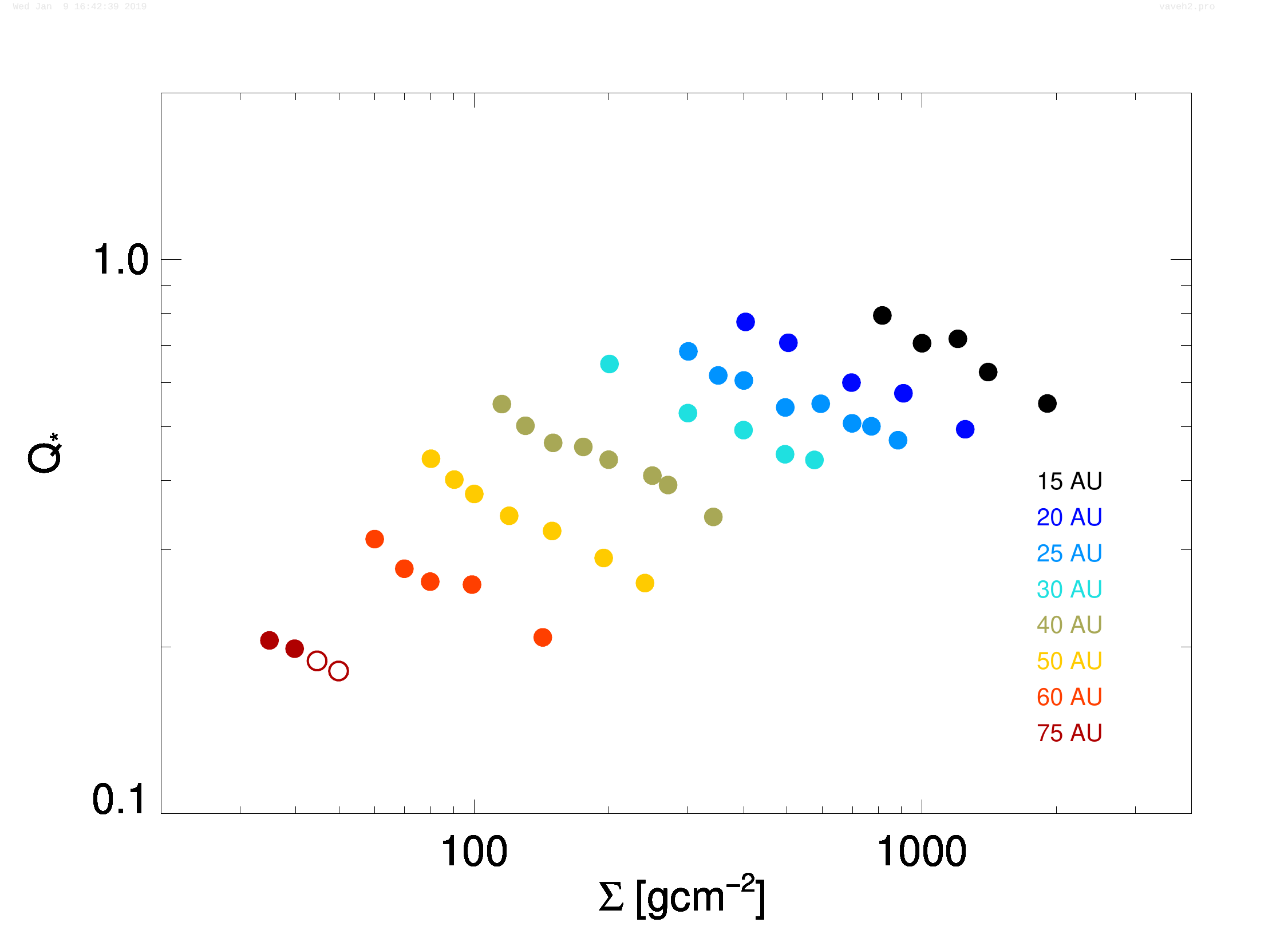} 
  \includegraphics[width=\columnwidth]{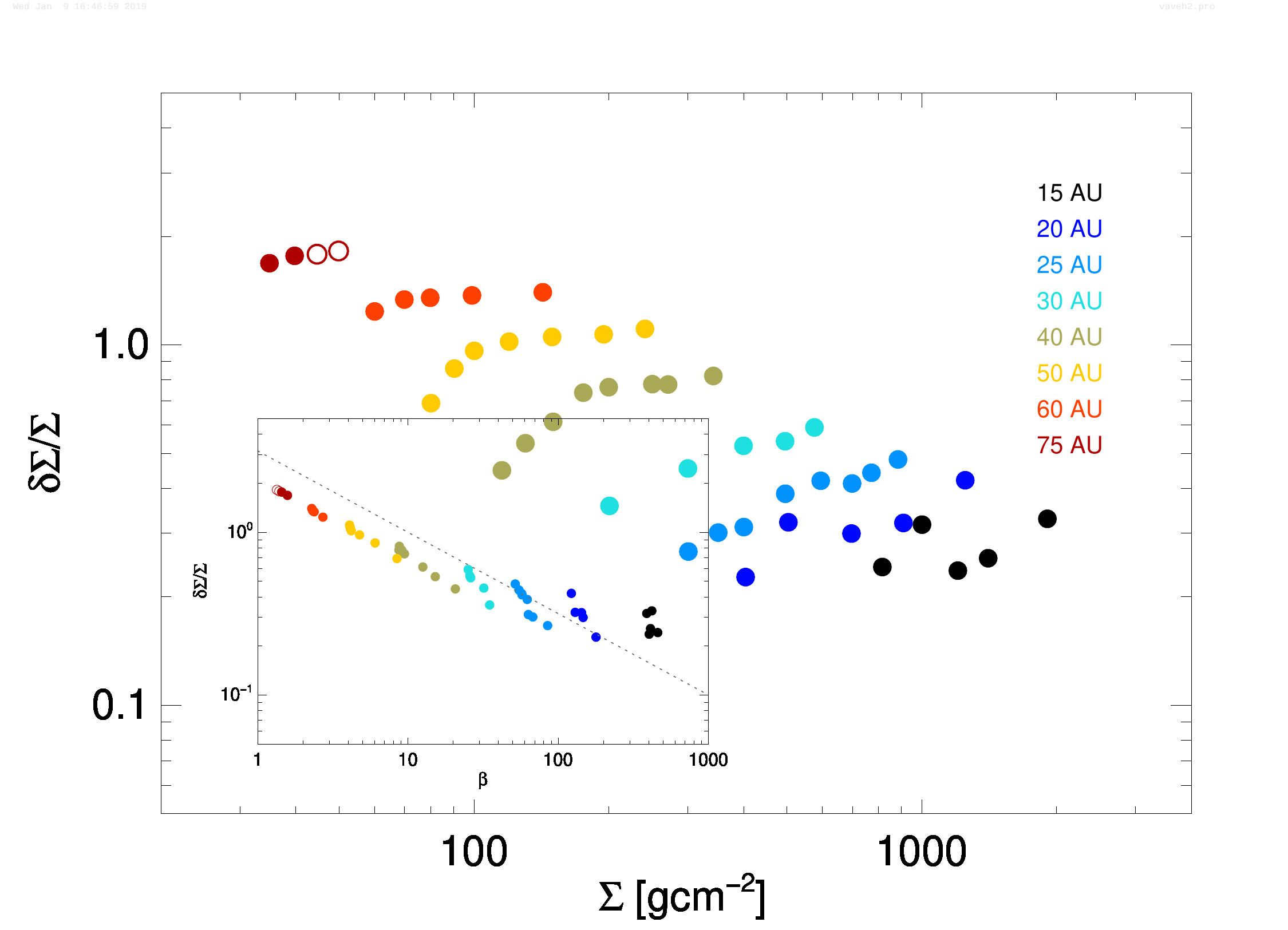} 
  \caption{Local Toomre's parameter $Q_\sm$ (upper) and the fractional density fluctuation $\delta\Sigma/\Sigma$ (lower).
    The usage of colours and symbols is the same as in Fig. \ref{fig:sigma_ave}.}
  \label{fig:sigma_min}
\end{figure}
\begin{figure*}
  \includegraphics[width=0.72\textwidth]{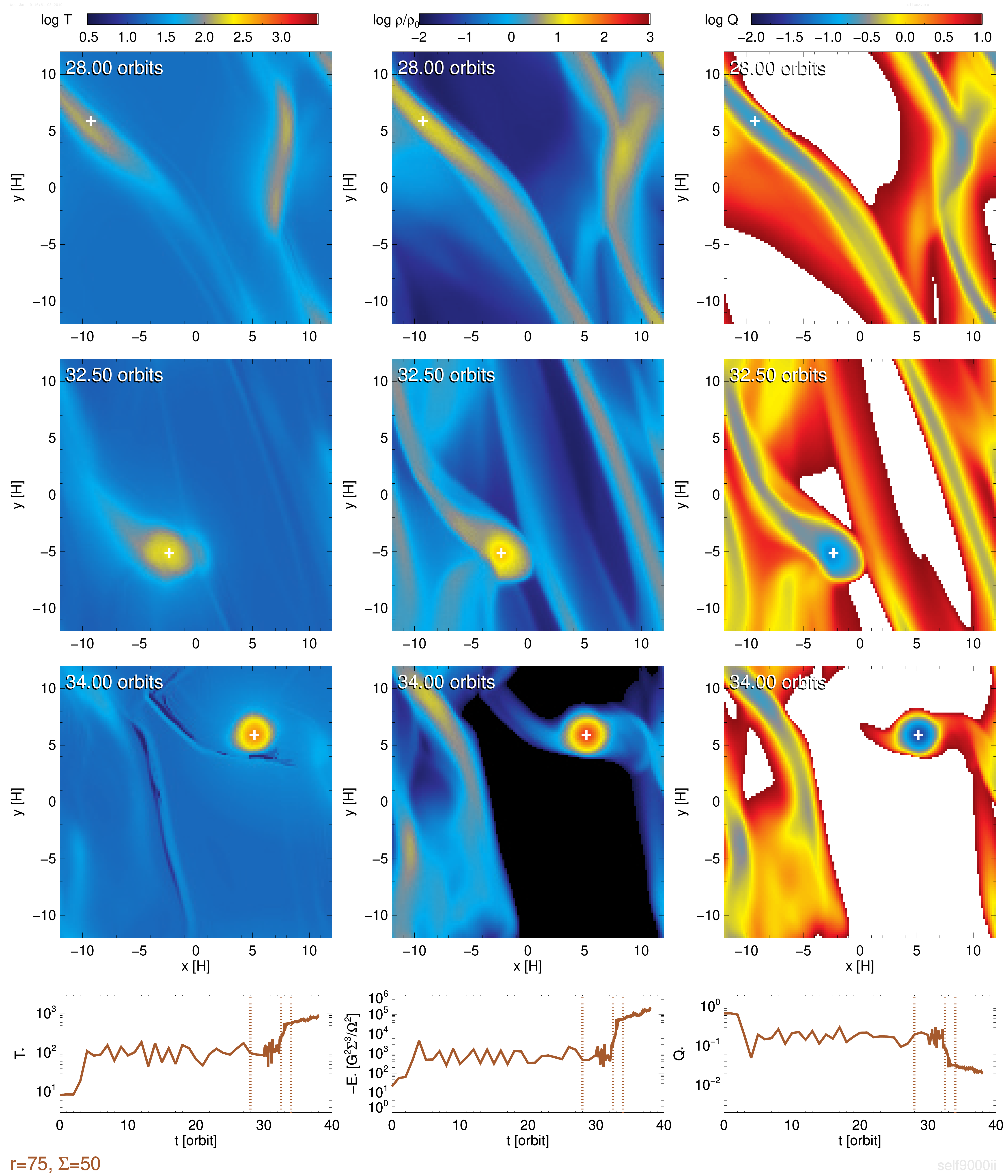} 
  \caption{Same as Fig. \ref{fig:self6004hh}, except for the transition case of $\Sigma = 50$ gcm$^{-2}$ at $r = 75$ AU, with the right column showing Toomre's parameter $Q$ instead of $\beta$.}
  \label{fig:self9000ii}
\end{figure*}

{\color{\modified}


As seen in Fig. \ref{fig:phase_diagram}, there are apparently two types of fragmentation boundaries. One is at $r \sim 75$ AU and the other is at $\Sigma \sim \Sigma_\text{0.2}$. 

Starting from \citet{Gammie:2001}, a consensus has been established that fragmentation occurs when the cooling time $\beta$ is less than a critical value of order unity. This is because a clump contracts to a bound object if the stochastic shock heating fails to catch up to the imposed cooling, which is especially expected when the cooling time $\beta$ is short \citep[e.g.][]{Paardekooper:2012}. As we discussed in Section \ref{sec:cooling_time}, the space-averaged cooling time scales as $\beta_\text{ave} \propto \Omega^3\Sigma^0$ and is as short as $\beta_\text{ave} \sim 1$ at $r = 75$ AU.
Therefore, the fragmentation boundary apparent at $r \sim 75$ AU in our simulations corresponds to the minimum cooling time ($\beta_\text{ave}\sim 1$) that can sustain the gravito-turbulence without fragmentation. Alternatively, because $\alpha \propto \beta_\text{ave}^{-1}$ (as we discussed in Section \ref{sec:stress_and_alpha}), the minimum cooling time can be redefined by the maximum $\alpha (\sim 1)$ that arises in the gravito-turbulence. 

A short cooling time in the gravito-turbulence has an important consequence. That is, the density fluctuation is expected to be anti-correlated to the averaged cooling time as
\begin{align}
  \frac{\delta\Sigma}{\Sigma} \propto \frac{1}{\sqrt{\beta_\text{ave}}},
\end{align}
which is derived from equating the cooling rate with the shock heating rate based on wave mechanics \citep{Cossins:2009, Rice:2011}. We see that this anti-correlation roughly holds in our results by comparing the top panel in Fig. \ref{fig:sigma_ave} and the lower panel in Fig. \ref{fig:sigma_min}, or as is explicitly shown in the inset of the lower panel of Fig. \ref{fig:sigma_min}, where the density fluctuation is computed as
\begin{align}
  &\frac{\delta\Sigma}{\Sigma} \equiv \frac{\int\sqrt{\left\langle \rho^2\right\rangle  - \left\langle \rho\right\rangle ^2}dz}{\int\left\langle \rho\right\rangle dz}.\label{eq:density_fluctuation}
\end{align}

As a consequence of the large density fluctuation, transient clumps in the gravito-turbulence have small $Q_\sm$ (see eq. \ref{eq:local_Q}). In our simulations, as $r$ increases, $\delta\Sigma/\Sigma$ increases and reaches values as large as $\sim 2$ at $r = 75$ AU (lower panel in Fig. \ref{fig:sigma_min}) whilst $Q_\sm$ decreases and reaches values as small as $\sim 0.2$ (upper panel in Fig. \ref{fig:sigma_min}).
With such small $Q_\sm$, a transient clump can be easily driven to a bound object by acquiring a small amount of mass via collision with other clumps or by accretion of ambient gas. Such fragmentation process is actually seen in the gravito-turbulence at $r = 75$ AU as shown in Fig. \ref{fig:self9000ii}. Note that the temperature of the clump does not decrease in the fragmentation process, which indicates that the fragmentation (i.e. reduction of $Q_\sm$) is caused by an increase in the local surface density, rather than by a decrease in the temperature due to cooling.

Because the cooling time $\beta_\text{ave}$ (or $\alpha$) does not depend on $\Sigma$, as shown in Figs. \ref{fig:sigma_ave} and \ref{fig:sigma_mdot}, the cooling time (or $\alpha$) cannot play a role in determining another fragmentation boundary at $\Sigma \sim \Sigma_{0.2}$ ($\Sigma$ corresponding to the initial Toomre's parameter $Q_0 = 0.2$). A simple explanation for this boundary is that the initial disc is too unstable against GI (i.e. $Q_0$ is too small). This is because the initial temperature is set as the radiative equilibrium temperature (eq. \ref{eq:T_0}), which does not depend on $\Sigma$. Therefore, the larger $\Sigma$ is, the smaller $Q_0$. However, the fragmentation boundary is not quite specific to that initial temperature. We have also evaluated cases in which the initial temperature is set so that $Q_0$ is kept at unity. The results did not change significantly because, as the simulation begins, the temperature quickly decreases to the radiative equilibrium temperature anyway owing to radiative cooling before GI develops. Therefore, when realistic radiative cooling is applied, a laminar disc of $\Sigma \gtrsim \Sigma_\text{0.2}$ would not evolve into the gravito-turbulence.
}

\subsection{Initial self-gravitating density waves}\label{sec:fragmentation_conditions}


\subsubsection{Stability of self-gravitating density waves}\label{sec:comparison_with_the_linear_stability_condition}
\begin{figure*}
  \includegraphics[width=0.72\textwidth]{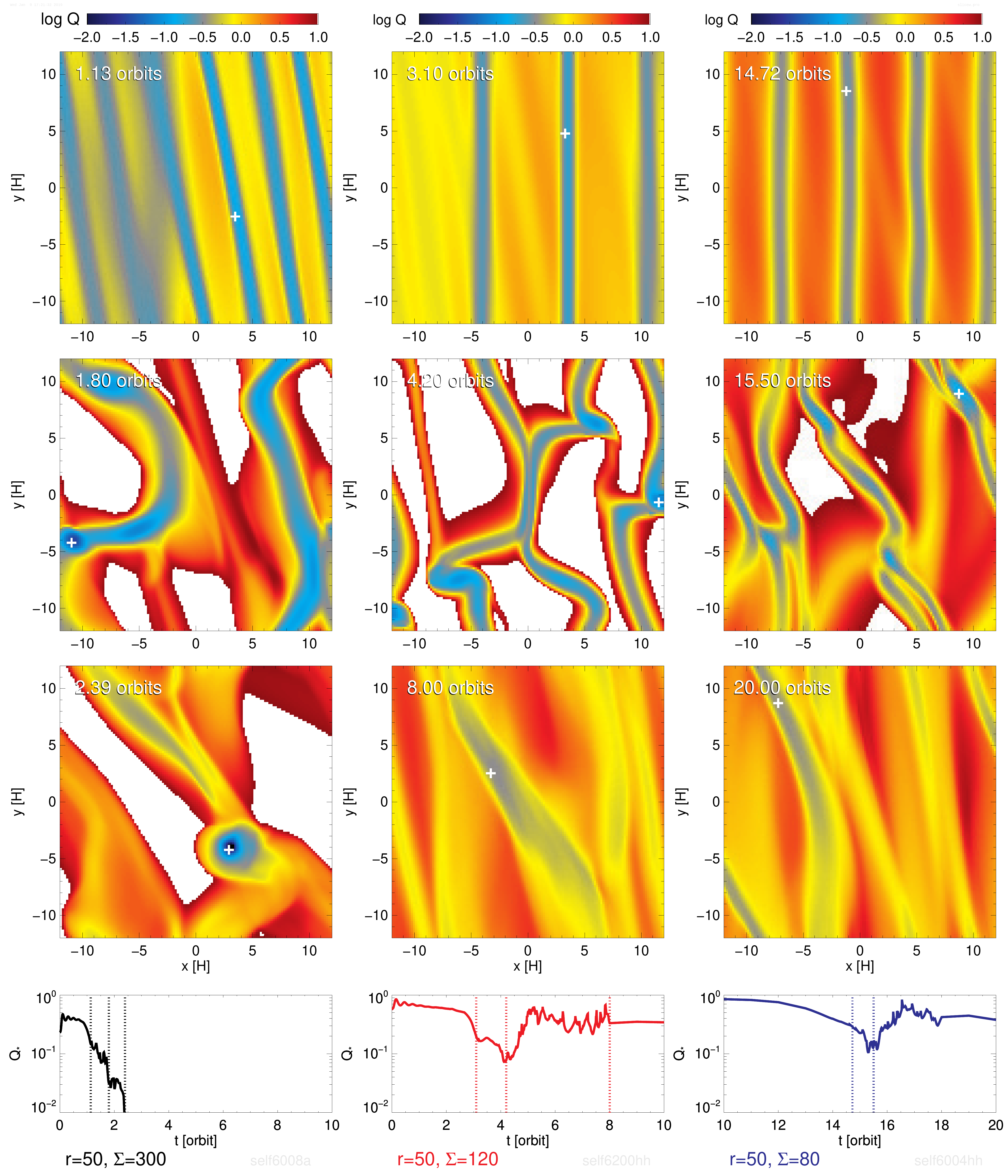} 
  \caption{Snapshots of $Q(x,y)$ in run R2 (left), R1 (middle) and R0 (right). In the bottom panels, the time evolution of $Q_\sm$ is shown. The cell of the minimum $E_\text{sg}$ is indicated as a white cross in the snapshots. The selected instances, indicated as vertical dotted lines in the bottom plots, are the same as those in Figs. \ref{fig:self6004hh}, \ref{fig:self6200hh}, and \ref{fig:self6008a}, from right to left.}
  \label{fig:slicew_hikaku_sigma}
\end{figure*}
\begin{figure*}
  \includegraphics[width=0.72\textwidth]{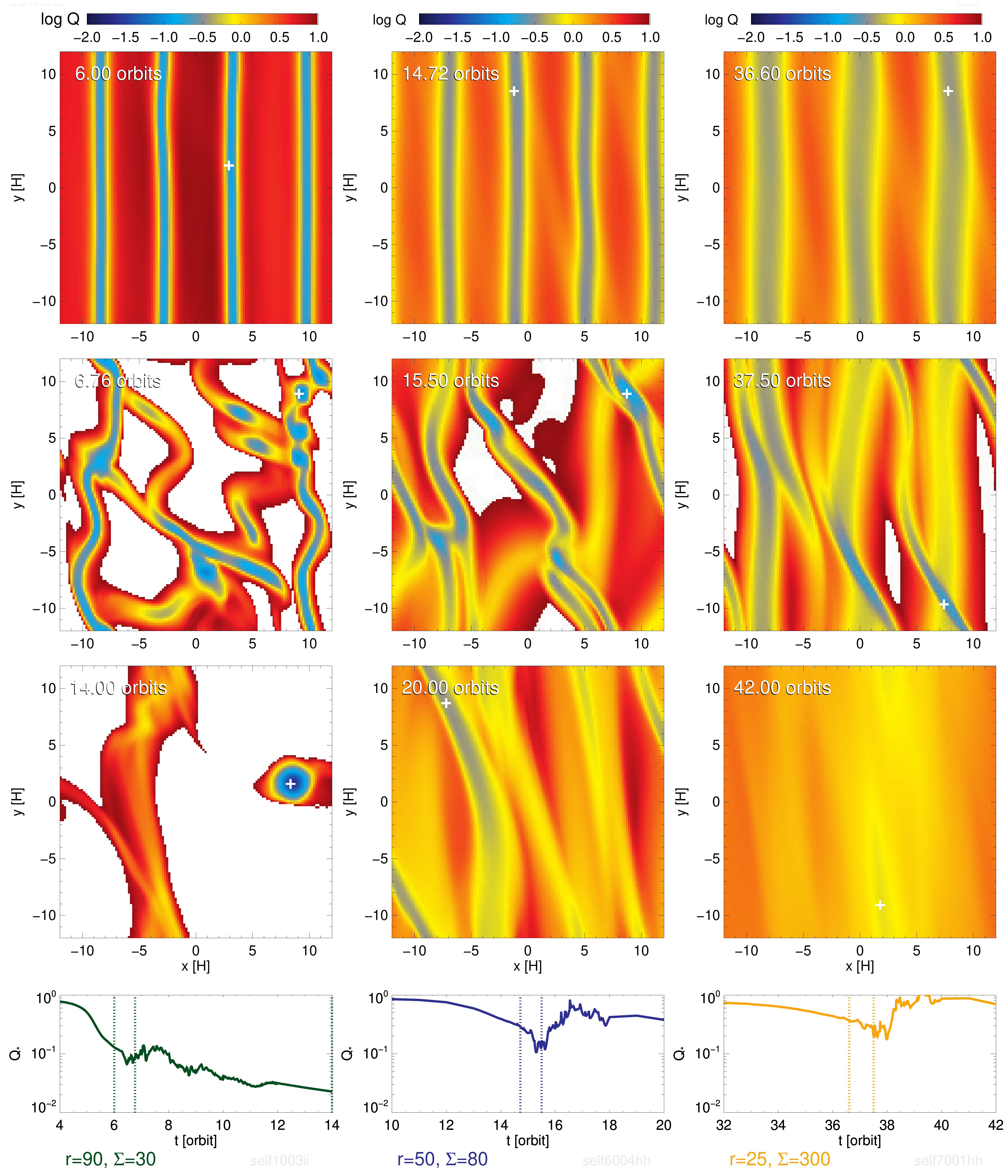} 
  \caption{Same as Fig. \ref{fig:slicew_hikaku_sigma}, except for run S2 (left), S1 (middle), and S0 (right). The selected three instances, indicated as vertical dotted lines in the bottom plots, are the same as those in Figs. \ref{fig:self7001hh}, \ref{fig:self6004hh}, and \ref{fig:self1003ii}, from right to left.}
  \label{fig:slicew_hikaku_radius}
\end{figure*}
{\color{\modified}

Here we examine again Fig. \ref{fig:condition_hikaku_q}, where we compare the time evolution of $Q_\sm$ amongst the runs listed in Table \ref{table:run_list}. In every run, after the initial plateau, there is a period of monotonic decrease in $Q_\sm$, which corresponds to nonlinear growth of the axisymmetric density waves.
As $Q_\sm$ decreases in time, the density waves become more strongly self-gravitating, and eventually become unstable when $Q_\sm$ decreases to a critical value $Q_\text{crit}$.
Note that the value of $Q_\text{crit}$ depends on $\Sigma$ and $r$; the larger $\Sigma$ or the larger $r$ is, the smaller $Q_\text{crit}$ becomes, which is also seen in the top rows of Figs. \ref{fig:slicew_hikaku_sigma} and \ref{fig:slicew_hikaku_radius}).


To understand the above stability of the initial self-gravitating density waves, we consult the linear analysis given in \citet{Takahashi:2016}. According to their analysis, a density wave (precisely, a two-dimensional density ring) of line mass $M_\text{L}$ and finite width $2W$ is unstable to non-axisymmetric perturbations when the Toomre's parameter of the density wave satisfies the following condition:
\begin{align}
  \tilde{Q} \equiv \frac{c_\text{s}(2\Omega)}{\pi GM_\text{L}/(2W)} < \tilde{Q}_\text{crit}(l),\label{eq:f}
\end{align}
where
\begin{align}
  & l \equiv \frac{(2W)}{c_\text{s}/(2\Omega)}\label{eq:l}
\end{align}
is the width of the density wave normalised by $c_\text{s}/(2\Omega)$.
Here, the tilde symbol of $\tilde{Q}$ denotes that it is evaluated with $\kappa = 2\Omega$ assuming that the density wave is rigidly rotating. The instability condition (\ref{eq:f}) states that the critical value $\tilde{Q}_\text{crit}$ depends on its normalised width $l$. Specifically, as shown in Fig. \ref{fig:condition_hikaku_ff}, as $l$ decreases, $\tilde{Q}_\text{crit}$ decreases.\footnote{The numerical data of $\tilde{Q}_\text{crit}(l)$ was kindly provided by Sanemichi Takahashi.} This is because the long-range effect of self-gravity to drive the non-axisymmetric instability is reduced in narrower waves and thus more surface density is required for instability. The decrease of $\tilde{Q}_\text{crit}$ is notable when $l \lesssim 1$, that is, when the width ($2W$) becomes comparable to or less than the scale height ($\propto c_\text{s}/(2\Omega)$). On the other hand, the critical value $\tilde{Q}_\text{crit}(l)$ becomes unity in the limit of infinite width ($l \to \infty$), which corresponds to the usual Toomre's condition, $Q < 1$.

To compare our simulation results with the linear stability condition, we evaluate the quantities $\tilde{Q}$ (eq. \ref{eq:f}) and $l$ (eq. \ref{eq:l}) for the density wave that contains the cell of the minimum self-gravitational energy, $(x_\sm,y_\sm,z_\sm=0)$. Specifically, we assume that the density wave is parallel to the $y$-axis (see the top rows of Figs. \ref{fig:slicew_hikaku_sigma} and \ref{fig:slicew_hikaku_radius}) and approximate the local surface density distribution across the density wave as Gaussian, $\sigma (x,y_\sm) \equiv \sigma_\text{max}e^{-((x-x_\sm)/\Delta x)^2/2}$. The width of a density wave of such Gaussian profile is somewhat arbitrary, but here we define $W \equiv 1.5\Delta x$ \citep[following][]{Takahashi:2016}, where the line mass is computed as $M_\text{L} \equiv \int_{-W}^{W}\sigma(x,y_\sm)dx \sim 1.44\sigma_\text{max}W$. 

\subsubsection{Comparison with linear analysis}
\begin{figure}
  \includegraphics[width=\columnwidth]{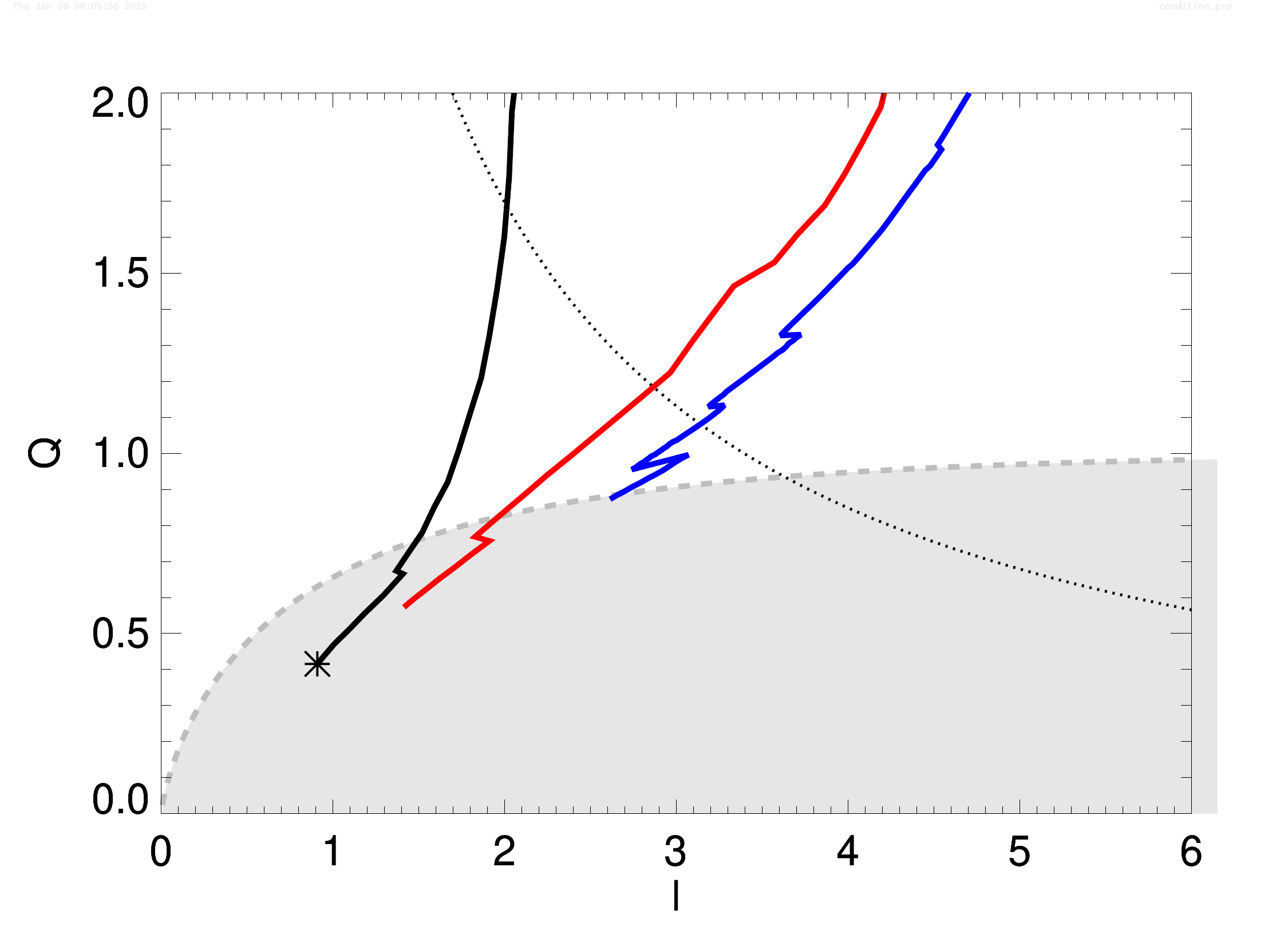} 
  \includegraphics[width=\columnwidth]{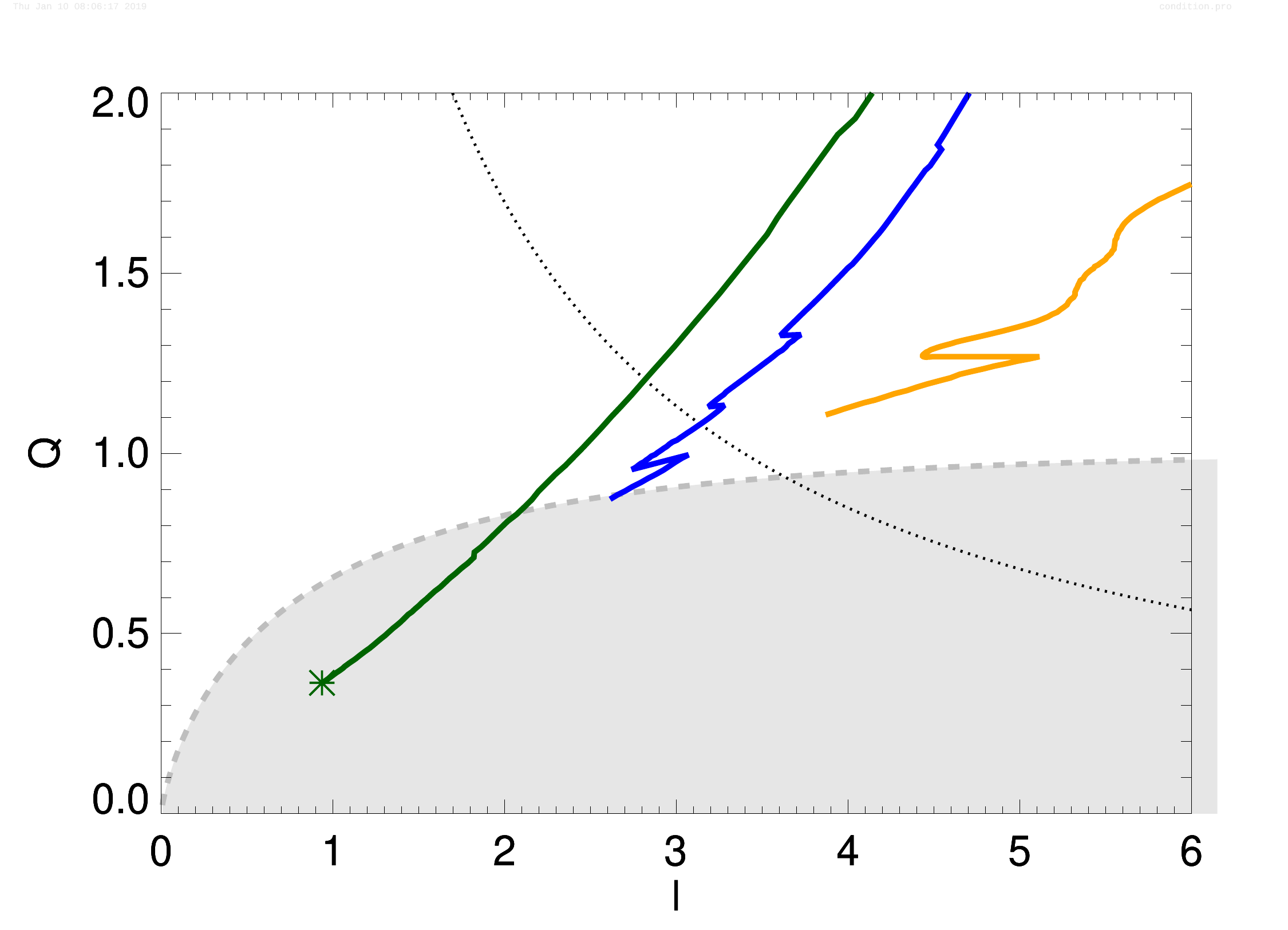} 
  \caption{Nonlinear growth of the axisymmetric density wave described in the $l$ -- $\tilde{Q}$ plane. The left edge of each curve corresponds to the epoch when the density wave becomes unstable. The asterisk at the left edge indicates that fragmentation occurs after the destabilisation of the density wave. The linear theory by \citet{Takahashi:2016} predicts that a density wave is unstable in the grey region. The dotted curve indicates the stability boundary determined by the Hill radius given by \citet{Rogers:2012}. The colour scheme is the same as for Fig. \ref{fig:condition_hikaku_q}. We note that because we are assuming that the epicyclic frequency $\kappa = 2\Omega$ in Section \ref{sec:comparison_with_the_linear_stability_condition}, the evaluated value of $\tilde{Q}$ here is twice that in other sections, where the epicyclic frequency is assumed as $\kappa = \Omega$.}
  \label{fig:condition_hikaku_ff}
\end{figure}
\begin{figure*}
  \includegraphics[width=0.72\textwidth]{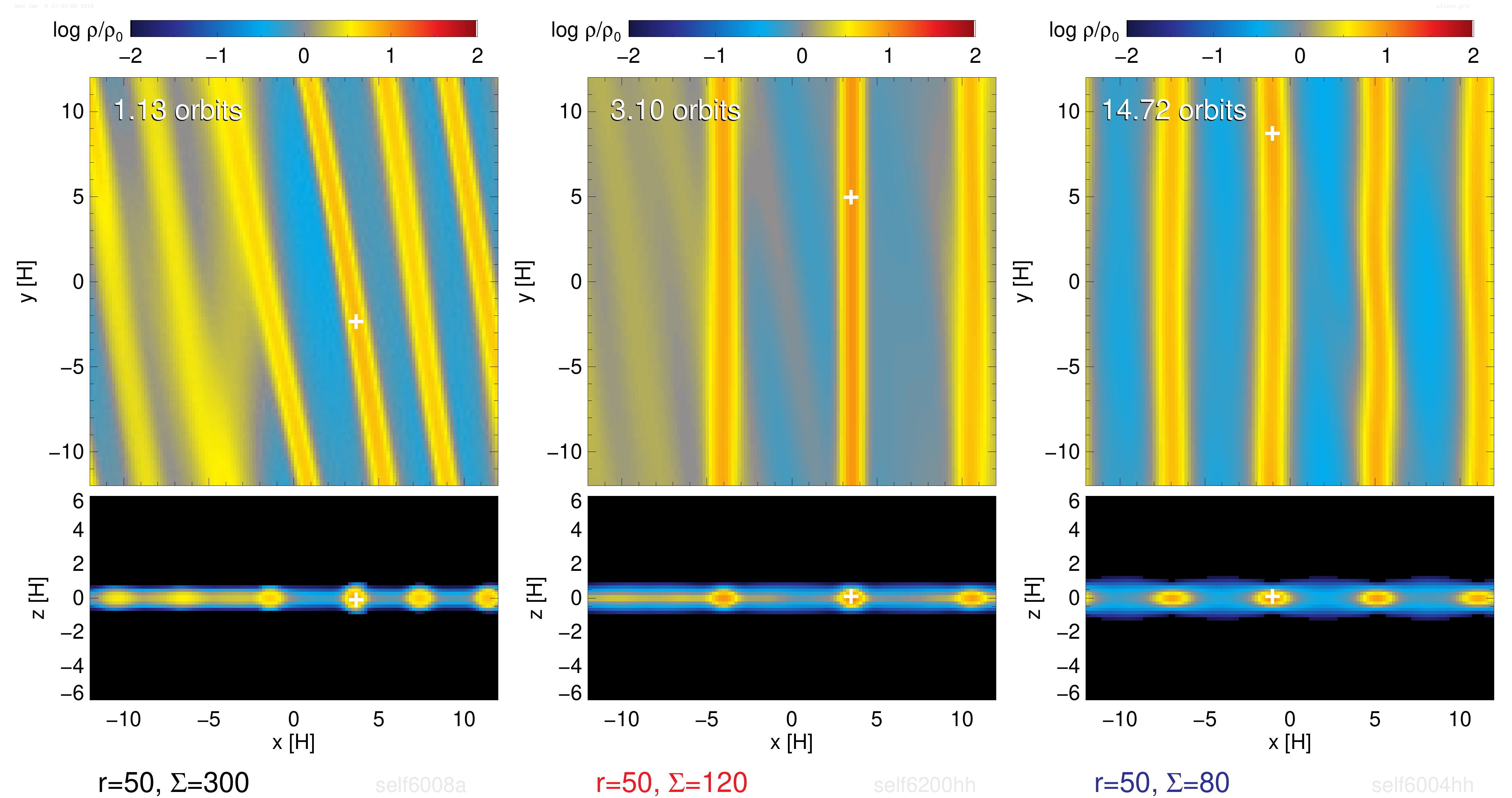} 
  \includegraphics[width=0.72\textwidth]{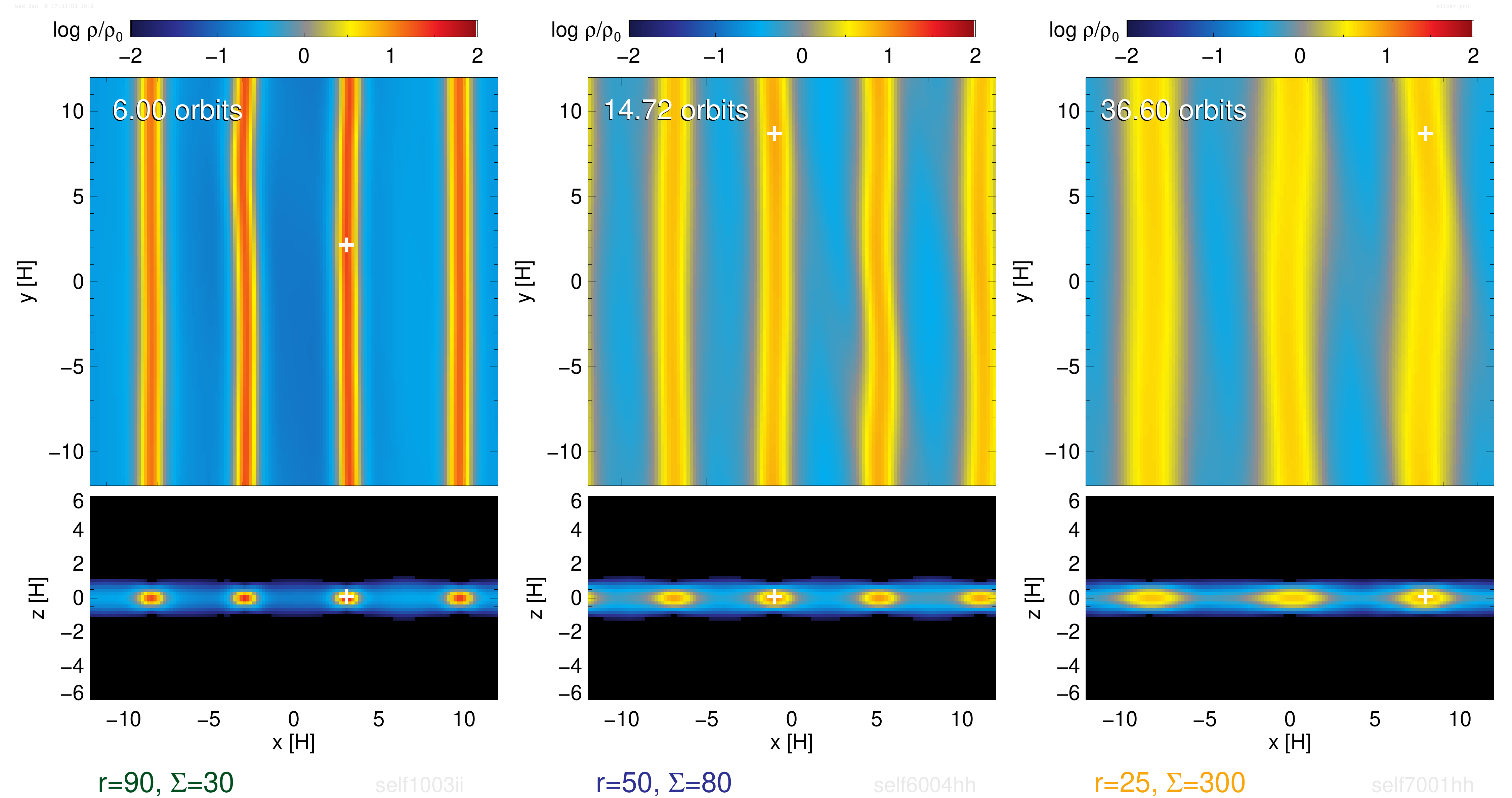} 
  \caption{Snapshots of density at the epoch when the axisymmetric density waves become unstable for the runs shown in Fig. \ref{fig:slicew_hikaku_sigma} (upstairs), and for runs shown in Fig. \ref{fig:slicew_hikaku_radius} (downstairs). In each pair of panels, the upper one shows $\rho(x,y,z=0)/\rho_0$ whilst the lower one shows $\rho(x,y=y_\sm,z)/\rho_0$. The cell of the minimum $E_\text{sg}$ is shown as a white cross in the snapshots. }
  \label{fig:slicev_hikaku}
\end{figure*}

In Fig. \ref{fig:condition_hikaku_ff}, the time evolution of the evaluated quantities $(l,\tilde{Q})$ is represented as a trajectory in the $l$-$\tilde{Q}$ plane for the five cases listed in Table \ref{table:run_list}. The density wave evolves from the upper right to the lower left by increasing its surface density ($\propto 1/\tilde{Q}$) as well as by decreasing its width ($\propto l$). Note that the left edge of each trajectory, which corresponds to the epoch when the density wave becomes unstable, is always found near the linear stability boundary, $\tilde{Q} = \tilde{Q}_\text{crit}(l)$. This indicates that the behaviour of the initial density waves in our simulations is approximately explained by the linear theory. (Perfect agreement is not expected as, for example, a single isolated density wave is assumed in the linear theory whilst this is not the case in our simulations.)

Fig. \ref{fig:condition_hikaku_ff} also shows that, as $\Sigma$ increases (upper panel) or $r$ increases (lower panel), the trajectory shifts leftward in the $l$-$\tilde{Q}$ plane. Increasing $r$ here also represents decreasing the cooling time, as shown in the bottom-right panels in Figs. \ref{fig:self7001hh}, \ref{fig:self6004hh}, and \ref{fig:self1003ii}. (Specifically, $\beta_\sm$ is $\sim 500$ at $r = 25$ AU, $\sim 30$ at $r = 50$ AU, and $\sim 6$ at $r = 90$ before the collapse.)
Therefore, when $\Sigma$ is larger or the cooling time is shorter at larger radii, the density waves grow narrower and thus the critical Toomre's parameter $\tilde{Q}_\text{crit}$ decreases. It is notable that regardless of changing $\Sigma$ or $r$, when the density waves become as narrow as $l \sim 1$, their collapse results in fragmentation.
Therefore, the two fragmentation boundaries ($\Sigma \sim \Sigma_{0.2}$ and $r\sim 75$ AU) may be translated to the single condition, $l \sim 1$, in terms of the width of the initial density waves \citep[c.f.][]{Tsukamoto:2015}. 

The dependence of the normalised width $l$ on $\Sigma$ or $r$ can also be observed in Fig. \ref{fig:slicev_hikaku}, where snapshots of density in the midplane $\rho(x,y,z=0)$ as well as in the $x$-$z$ plane $\rho(x,y=y_\sm,z)$, at the epoch when the density waves become unstable, are compared amongst the five runs. The normalised width $l = 2W/(c_\text{s}/2\Omega)$ is regarded as the ratio of the width to the thickness of the density wave (except for a numerical factor of order unity). Then, we see clearly that as $\Sigma$ increases or $r$ increases, the density waves become narrower in terms of their thickness.

}

\subsection{Transition from fragmentation to runaway collapse}\label{sec:transitions2}
\begin{figure}
  \includegraphics[width=\columnwidth]{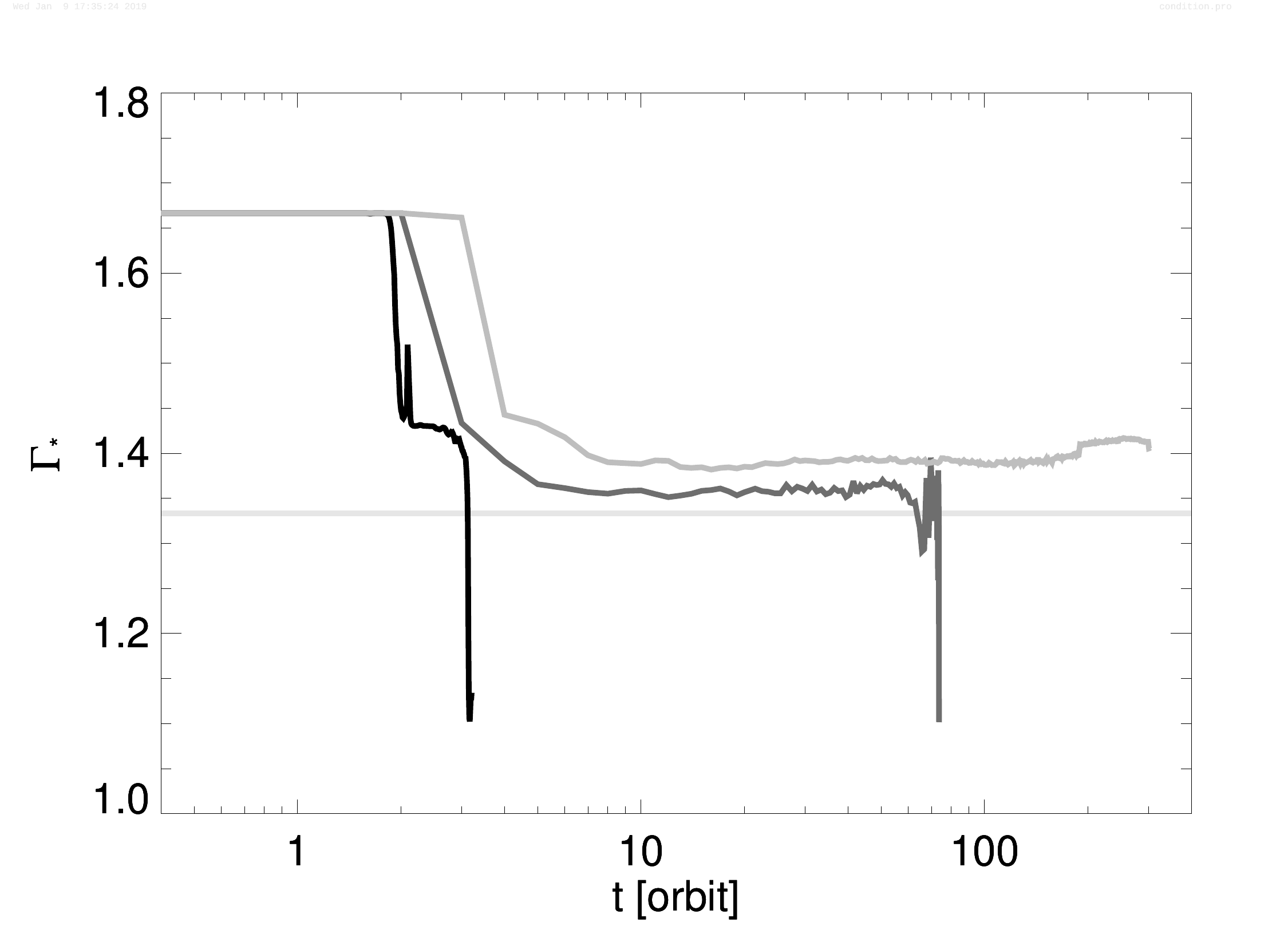} 
  \caption{Same as Fig. \ref{fig:condition_hikaku_gamma}, except it compares cases for three different values of $\Sigma$ at $r = 75$ AU: $\Sigma = 55$ (grey), $70$ (dark grey), and $80$ gcm$^{-2}$ (black).}
  \label{fig:condition_transient_gamma}
\end{figure}

In some cases at $r = 75$ and $90$ AU, a transition from fragmentation to runaway collapse was observed. In Fig. \ref{fig:condition_transient_gamma}, we compare the time evolution of $\Gamma_\sm$ in three cases involving different outcomes at $r = 75$ AU, which are a fragmentation case ($\Sigma = 55$ gcm$^{-2}$), a transition case ($\Sigma = 70$ gcm$^{-2}$), and a runaway collapse case ($\Sigma = 80$ gcm$^{-2}$).

In the smallest $\Sigma$ case, a pressure-supported clump was formed and survived a few hundreds of orbits, where the core temperature $T_\sm$ was maintained at the value that corresponds to $\Gamma_\sm \sim 1.4$. In the intermediate $\Sigma$ case, a pressure-supported clump was also formed, but $T_\sm$ was larger (and thus $\Gamma_\sm$ was smaller) owing to stronger self-gravity. Furthermore, because the clump was not completely isolated, it accreted mass from the ambient medium and $T_\sm$ rose accordingly. Eventually, at $t \sim 73$ orbits, $T_\sm$ exceeded the H$_2$ dissociation temperature to cause $\Gamma_\sm < 4/3$, leading to runaway collapse. In the largest $\Sigma$ case, owing to the strongest self-gravity, $T_\sm$ quickly exceeded the H$_2$ dissociation temperature and thus $\Gamma_\sm$ became smaller than $4/3$ just after fragmentation occurred.

In summary, because a pressure-supported clump formed in fragmentation is not equilibrated with the ambient medium, it inevitably evolves by accretion or radiative cooling. Especially, when the core temperature of a formed clump is close to the H$_2$ dissociation temperature, runaway collapse may eventually occur as a result of such evolution.

\section{Discussion}

\subsection{Comparison with previous studies}
{\color{\modified}
\subsubsection{Fragmentation boundaries in phase diagrams}
In this section, we compare our phase diagram of the nonlinear outcome of GI (Fig. \ref{fig:phase_diagram}) with two previous studies, \citet{Johnson:2003} and \citet{Clarke:2009}.

More direct comparison is possible with \citet{Johnson:2003}, who performed 2D local shearing box simulations with a cooling function based on a one-zone model of optically thick disks. Unlike our simulations, a simple EOS was used and irradiation was not taken into account in their simulations. In Fig. \ref{fig:johnson03}, we plot our results on the $\Sigma$-$\Omega$ plane for direct comparison with their Fig.7. The two solid curves drawn are taken from their Fig. 7, where the lower curve connects runs that show no signs of fragmentation whilst the upper curve connects those showing definite fragmentation. 
When compared over a common range of $\Omega$, the fragmentation boundary is qualitatively similar. That is, at every $\Omega$, a critical $\Sigma$ exists beyond which fragmentation occurs, although our critical $\Sigma$ is consistently slightly larger. On the other hand, the interpretation of what causes the fragmentation differs. They attribute the fragmentation to short space-averaged cooling time. In contrast, we did not see such $\Sigma$ dependence of the cooling time in our simulations (Fig. \ref{fig:sigma_ave}). Rather, we attribute the fragmentation at $\Sigma\sim\Sigma_{0.2}$ to small values of $Q_0$, as discussed in Section \ref{sec:fragmentation_condition}.


As we discussed in Section \ref{sec:stress_and_alpha}, when gravito-turbulence is established, the mass accretion rate $\dot{M}$ can be evaluated from the vertically-integrated stress $\langle W_{xy}\rangle$ (eq. \ref{eq:mdot_stress}), and we found a unique positive correlation between $\dot{M}$ and $\Sigma$ (lower panel in Fig. \ref{fig:sigma_mdot}). Using these results, we can compare our Fig. \ref{fig:phase_diagram} with Fig. 4 in \citet{Clarke:2009}, who analytically obtained the gravito-turbulence solutions assuming $Q=1$ and the local thermal and hydrostatic equilibrium. They utilised the maximum $\alpha$ value of $0.06$ in the gravito-turbulence to identify the fragmentation boundary, which was found at $70$ AU. As compared in Fig. \ref{fig:clarke09}, the location of their fragmentation boundary at $r = 70$ AU is fairly close to the one found in our simulations.
On the other hand, in \cite{Clarke:2009}, the $\alpha$ value increases with the accretion rate at high mass accretion rates and thus a fragmentation boundary also exists at the high mass accretion rate side \citep[see also][]{Zhu:2012,Forgan:2013}. In our simulations, although there also exists a fragmentation boundary at the high mass accretion rate side, it is not due to the dependence of $\alpha$ on the mass accretion rate, but is again due to the initial small $Q_0$, as discussed in the above.


\begin{figure}
  \includegraphics[width=\columnwidth]{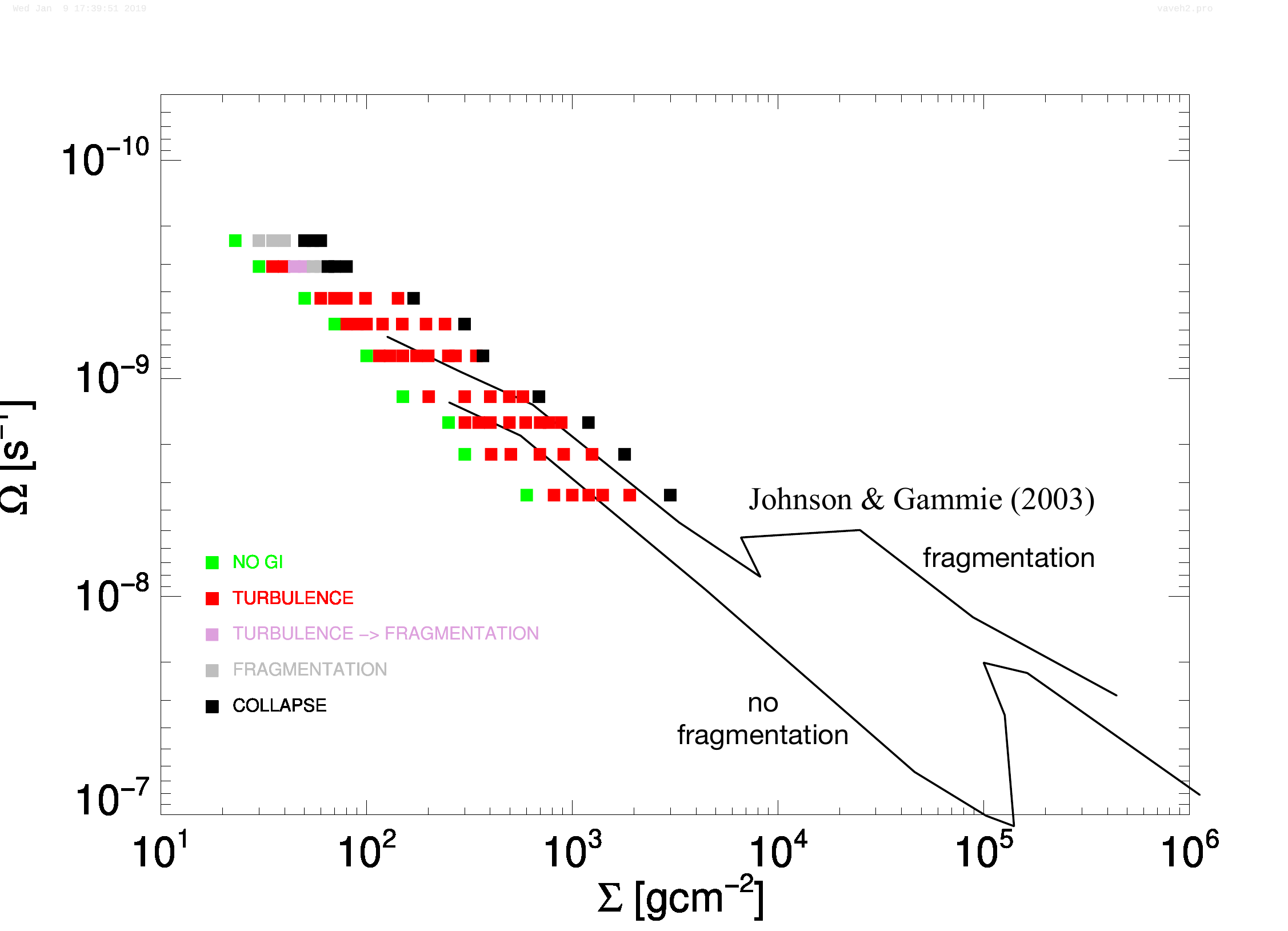} 
  \caption{Comparison with \citet{Johnson:2003}. The nonlinear outcomes of GI shown in Fig. \ref{fig:phase_diagram} are replotted in the same frame as their Fig. 7 (the vertical and horizontal axes are exchanged). The two black curves are their ``critical curves''.} 
  \label{fig:johnson03}
\end{figure}

\begin{figure}
  \includegraphics[width=\columnwidth]{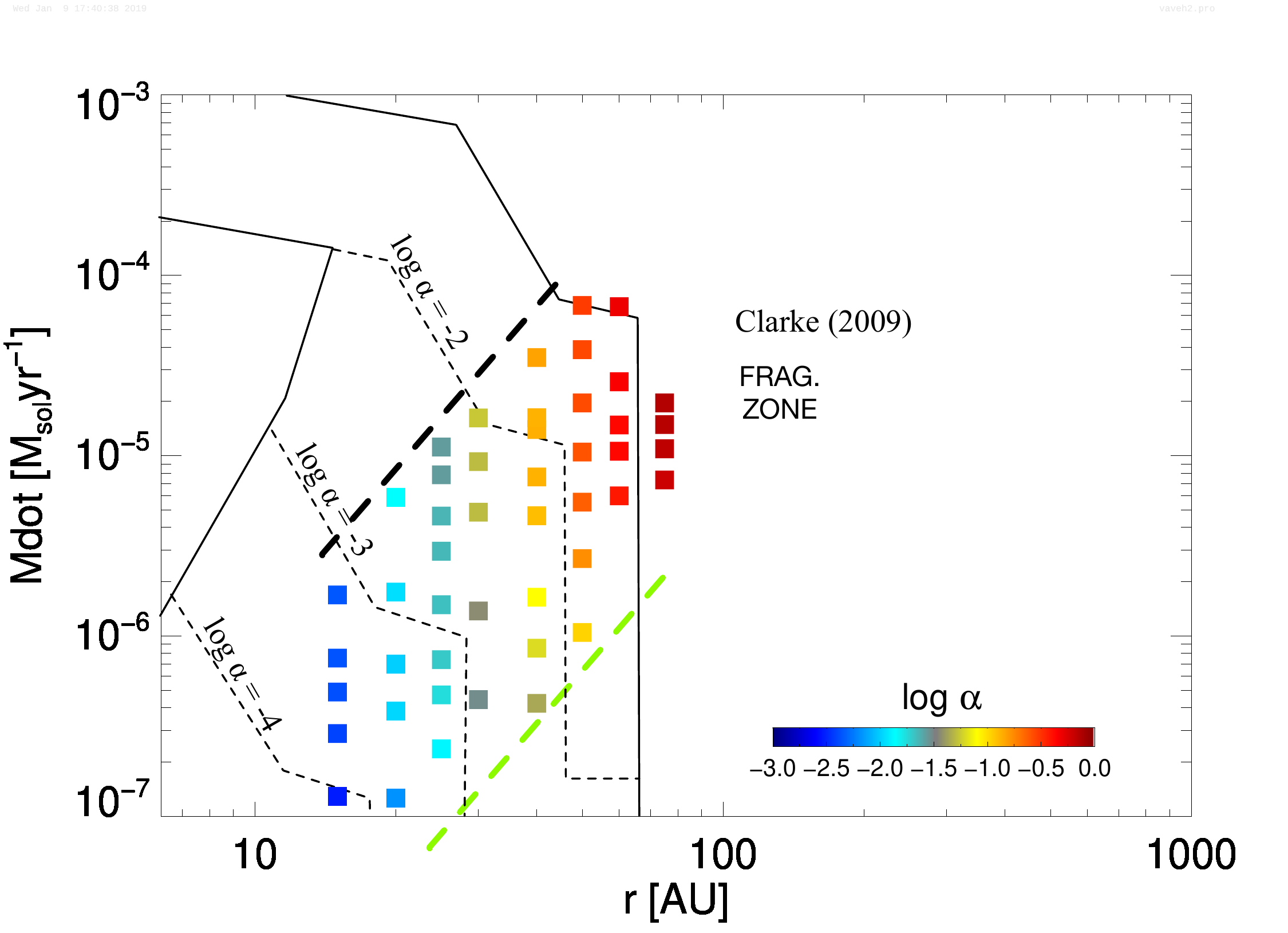} 
  \caption{Comparison with \citet{Clarke:2009}. The colour shows the $\alpha$ value of the gravito-turbulence solutions in our simulations. In \citet{Clarke:2009}, the gravito-turbulence solutions exist in the region between the two black solid curves with the $\alpha$ value indicated by the black dotted curves. 
    The black and green dashed lines roughly denote $\Sigma = \Sigma_{0.2}$ and $\Sigma = \Sigma_1$, respectively.}
  \label{fig:clarke09}
\end{figure}


\subsubsection{Stability of density waves and fragmentation}
As we discussed in Section \ref{sec:fragmentation_conditions}, the correlation between the width of the initial density waves and the critical Toomre's parameter $Q_\text{crit}$ found in our simulations is mostly consistent with the linear stability analysis presented by \citet{Takahashi:2016} (see Fig. \ref{fig:condition_hikaku_ff}). On the other hand, there is a notable difference between our fragmentation condition and theirs, which apparently comes from the difference between their global and our local simulations. They claimed that fragmentation occurs if and only if spiral density waves become unstable against the non-axisymmetric instability. Namely, the fragmentation condition is identical to the instability condition of the density waves (c.f. their Fig. 1). 
In contrast, in our simulations, the initial axisymmetric density waves always became unstable and collapsed. However, it is only when the density waves grow as narrow as $\tilde{Q}_\text{crit}(l = 1) \sim 0.4$ that the collapse results in fragmentation, which takes place either when $\Sigma > \Sigma_{0.2}$ or when $r > 75$ AU. 

Our conclusion that fragmentation occurred when the density waves became narrow enough appears to be similar to the fragmentation condition determined by the Hill radius proposed by \citet{Rogers:2012}. Their fragmentation condition can be rewritten as \citep[c.f.][]{Takahashi:2016},
\begin{align}
  Q < Q_\text{crit}(l) \equiv \frac{32}{3\pi}\frac{1}{l},\label{eq:rogers}
\end{align}
which is also plotted in Fig. \ref{fig:condition_hikaku_ff}. As shown, the initial axisymmetric self-gravitating density waves in our simulations kept growing without fragmentation even after they entered the above unstable region (eq. \ref{eq:rogers}). Therefore, our simulation results may not be explained in terms of the Hill radius as proposed by \citet{Rogers:2012}.

\subsubsection{Steady accretion driven by gravito-turbulence}\label{sec:steady_accretion}
\begin{figure}
  \includegraphics[width=\columnwidth]{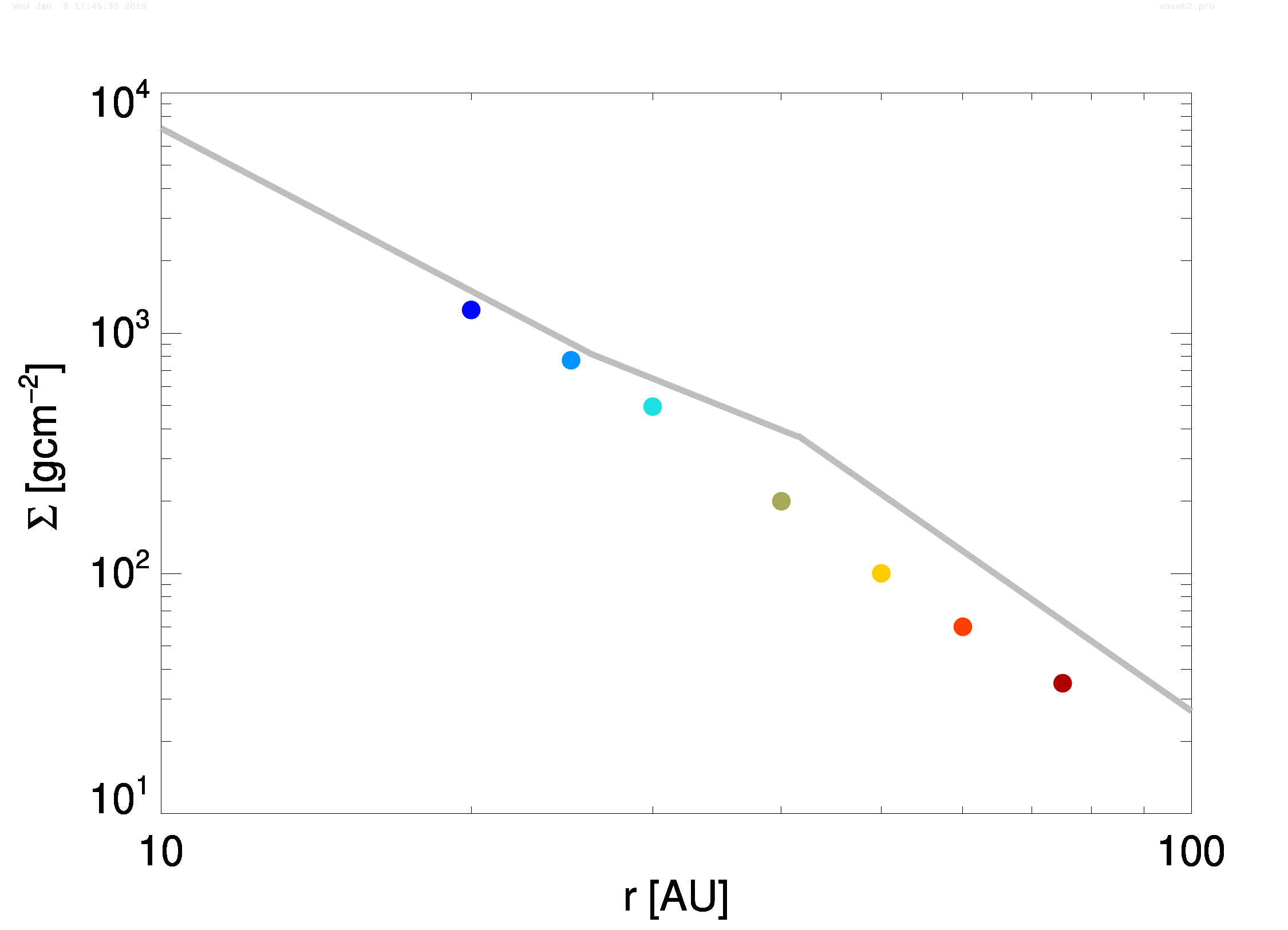} 
  \includegraphics[width=\columnwidth]{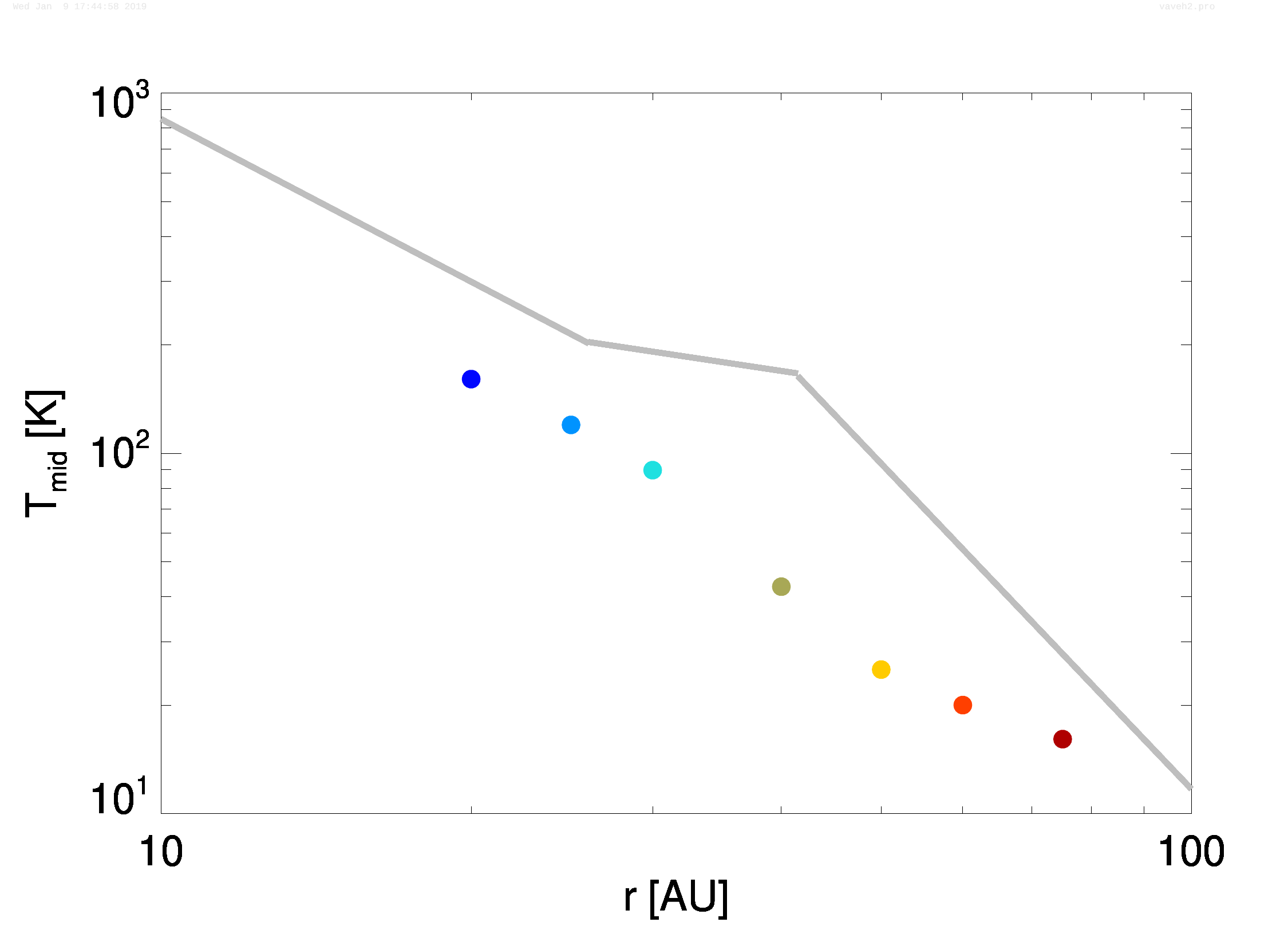} 
  \includegraphics[width=\columnwidth]{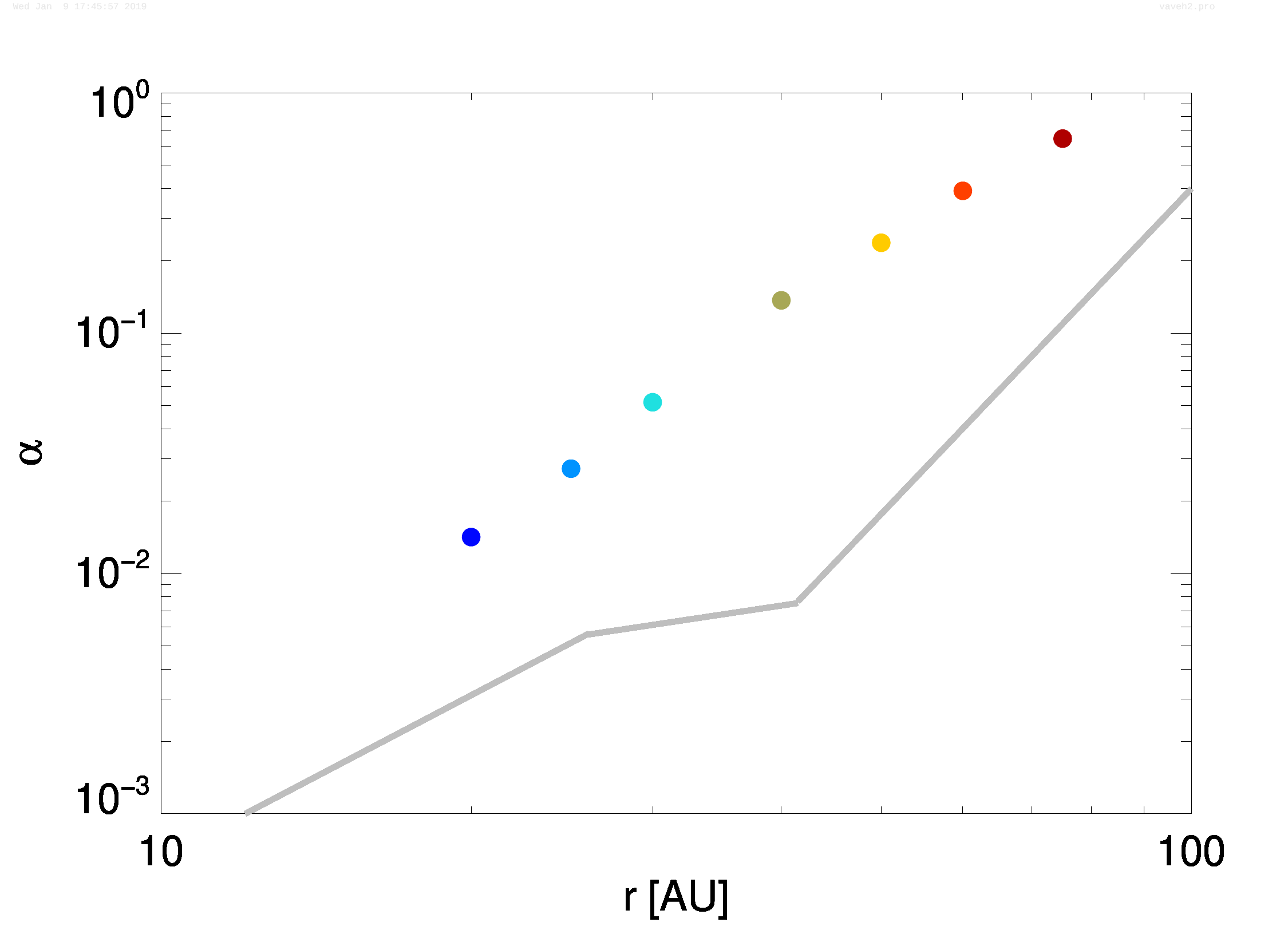} 
  \caption{Profiles of $\Sigma$ (top), $T_\text{mid}$ (middle), and $\alpha$ (bottom) of a steady accretion disc of $\dot{M}=7\times10^{-6}$ $M_\odot$yr$^{-1}$ based on our simulations. The grey curve in each panel shows the profile in the analytical model of \citet{Clarke:2009} at the same mass accretion rate.}
  \label{fig:steady_hikaku}
\end{figure}

Next we examine the gravito-turbulence in thermal equilibrium. Fig. \ref{fig:sigma_mdot} shows that steady accretion driven solely by gravito-turbulence is possible for a range of radii, which depends on the mass accretion rate \citep[c.f.][]{Rice:2009}.
For example, mass accretion of $\dot{M} = 7\times10^{-5}$ $M_\odot$yr$^{-1}$ would be possible for $20 \lesssim r \lesssim 75$ AU. To construct a steady accretion disc for that accretion rate, we pick up a solution of $\dot{M} \sim 7\times10^{-5}$ $M_\odot$yr$^{-1}$ at each radius between 20 and 75 AU and combine them. The radial profiles of $\Sigma$, $T_\text{mid}$, and $\alpha$ for such a steady accretion disc based on our simulations are plotted in Fig. \ref{fig:steady_hikaku}.

These profiles are directly compared with those of the analytical model by \citet{Clarke:2009} because they assume a central star of $1M_\odot$ as we did. In Fig. \ref{fig:steady_hikaku}, their profiles at $\dot{M} = 7\times10^{-6}$ $M_\odot$yr$^{-1}$ are also plotted. Among the three quantities, it is notable that the midplane temperatures are consistently higher in \citet{Clarke:2009}, indicating that radiative cooling is less effective in their case. This may be because they adopted the midplane opacity for cooler upper layers, which would overestimate the optical thickness of the disk because $\kappa \propto T^2$ (below the ice melt temperature). If this is the case, they would also overestimate the cooling time $\beta$, which leads to underestimating the $\alpha$ value ($\propto 1/\beta$ in thermal equilibrium) and then to overestimating $\Sigma$. These naturally explain the discrepancies between \citet{Clarke:2009}'s model and ours seen in Fig. \ref{fig:steady_hikaku}, although both models use a similar opacity (see the bottom panel in Fig. \ref{fig:sigma_ave}). Therefore, resolving the vertical structure with appropriate radiative transport is essential in determining the radial profile of the disc.


}
\subsection{Formation of bound clumps via gravitational instability}
{\color{\modified}
  Using 3D global disc simulations adopting the $\beta$ cooling, \citet{Boss:2017} showed that low $Q_0$ discs can fragment for high $\beta$ whilst high $Q_0$ discs can be stable for small $\beta$, which indicates the equal importance of the initial Toomre's parameter to the cooling time for fragmentation. Our simulations qualitatively agree with their results regarding the importance of the initial Toomre's parameter. Namely, Fig. \ref{fig:phase_diagram} shows that fragmentation is possible at any radius, or at any cooling time, provided that the surface density is as large as $\Sigma_\text{crit} \sim \Sigma_{0.2}$, or the initial Toomre's parameter is as small as $0.2$. On the other hand, beyond $r \sim 90$ AU, the critical surface density is relaxed to $\Sigma_\text{crit} \sim \Sigma_{1}$. That is, when the cooling time is as short as $\beta_\text{ave} \lesssim 1$, fragmentation always occurs at any value of $Q_0$ less than unity.
  
  Using the value of $\Sigma_\text{crit}$, we make a crude estimate of the mass of a bound clump formed in the collapse of the initial density waves as per \citep[c.f.][]{Rafikov:2005} 
\begin{align}
  & M_\text{clump} \sim (f H)^2 \Sigma_\text{crit} \sim \left(f\frac{{c_\text{s}}_0}{\Omega}\right)^2\left(\frac{{c_\text{s}}_0\Omega}{\pi G Q_0}\right),
\end{align}
which can be written in terms of the Jupiter mass $M_\text{J} = 2\times10^{30}$ g as
\begin{align}
  \frac{M_\text{clump}}{M_\text{J}} \sim
  \begin{cases}
    5 \left(\dfrac{f}{6}\right)^2\left(\dfrac{Q_0}{0.2}\right)^{-1}\left(\dfrac{r}{25\text{AU}}\right)^\frac{3}{4} &(r \lesssim 75\text{ AU}),\\
    3 \left(\dfrac{f}{6}\right)^2\left(\dfrac{Q_0}{1}\right)^{-1}\left(\dfrac{r}{90\text{AU}}\right)^\frac{3}{4}  &(r \gtrsim 90\text{ AU}).
  \end{cases}
\end{align}
The numerical factor $f$ here stands for the size of such a clump in terms of the scale height $H$, for which we employ a value of $\sim 6$ based on our simulations.
The above equation indicates that the minimum mass of a bound clump formed in the non-axisymmetric instability is several to ten times $M_\text{J}$ for the radii we have explored. 

So far as we have investigated, a pressure-supported clump once formed was never dissolved by the velocity shear, either surviving or being followed by runaway collapse. This indicates that the realistic cooling is so efficient that a formed clump remains compact enough to resist velocity shear.

}

\subsection{Dependence on the box size and the spacial resolution for fragmentation cases}\label{sec:dependence_on_numerical_issues}
In our simulations shown above, the box size and the spacial resolution were fixed as, respectively, $(L_x,L_y,L_z) = (24H, 24H, 12H)$ and $(N_x,N_y,N_z) = (128, 128, 64)$. They are the same as those used in the fiducial run in Paper I, where we showed that the results do not strongly depend on them when gravito-turbulence is established. Here we discuss how the results could depend on the box size or the spacial resolution when fragmentation occurs (i.e. $Q_\sm < 1$ in the final state). 

\begin{figure}
  \includegraphics[width=\columnwidth]{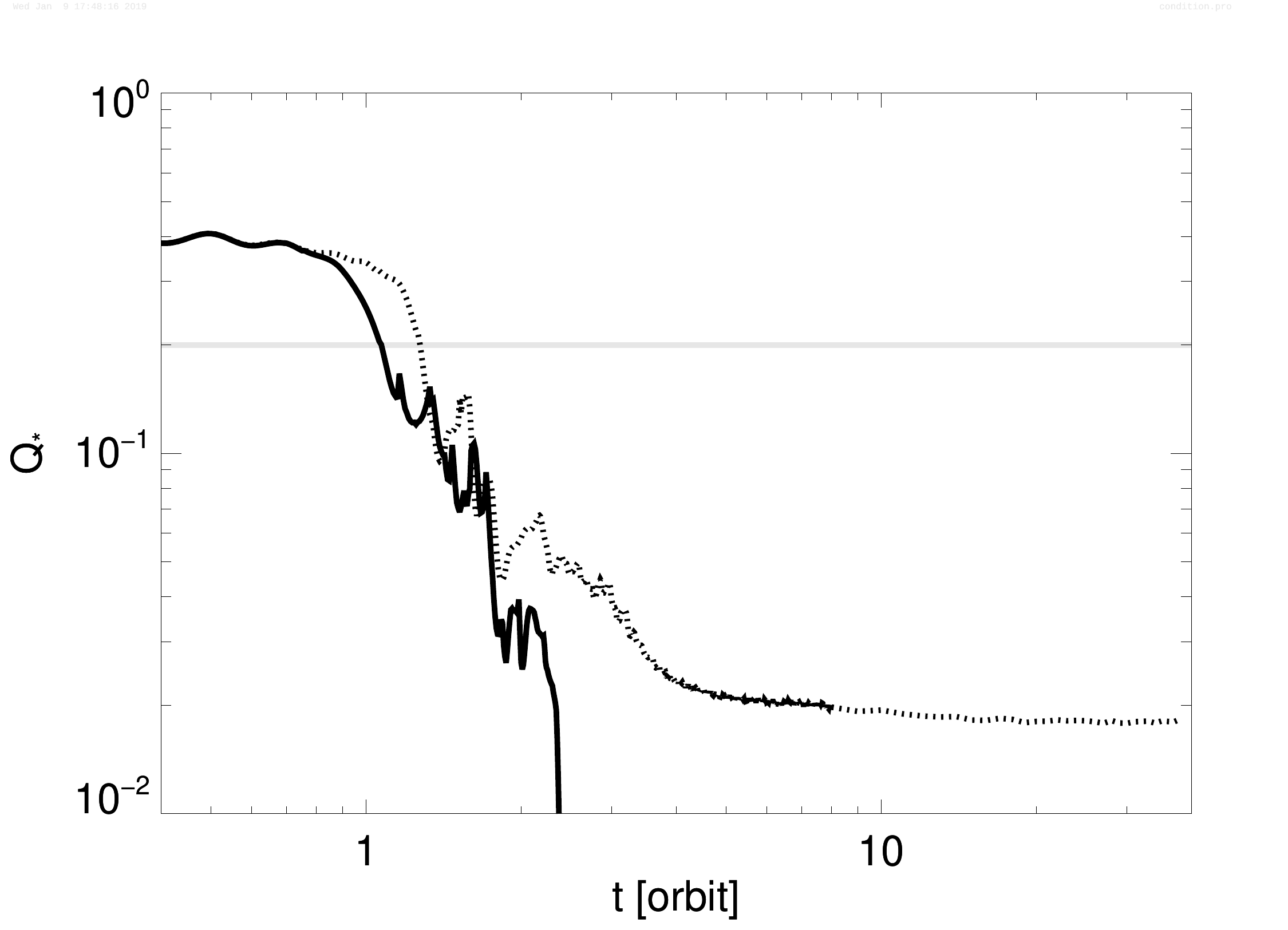} 
  \caption{Same as Fig. \ref{fig:condition_hikaku_q}, except for run R2 (thick) and the same run as R2 except using a simulation box of halved horizontal size (dotted).}
  \label{fig:condition_r50_box_q}
\end{figure}

Firstly we examine the box size dependence. Fig. \ref{fig:condition_r50_box_q} compares the time evolution of $Q_\sm$ in the case of $\Sigma = 300$ g cm$^{-3}$ at $r = 50$ AU (run S2) as well as that in the same case but with a halved box size, i.e. $(L_x,L_y,L_z) = (12H, 12H, 6H)$. In the case of the standard box size, $Q_\sm$ decreased below the critical value of $\sim 0.2$ and fragmentation occurred, which was followed by runaway collapse at $t \sim 2.3$ orbits. On the other hand, in the case of the halved box, although fragmentation occurred similarly, it was not followed by runaway collapse, and a pressure-supported clump survived instead. This means that mass concentration by self-gravity in the halved box, which contained one fourth the amount of mass contained in the standard box, was not enough to raise the core temperature of the clump above the H$_2$ dissociation temperature. Therefore, although we may always expect fragmentation beyond some critical $\Sigma$ at a given radius, whether runaway collapse follows the fragmentation depends on how much mass concentrates by self-gravity, which then may depend on the box size. 

\begin{figure}
  \includegraphics[width=\columnwidth]{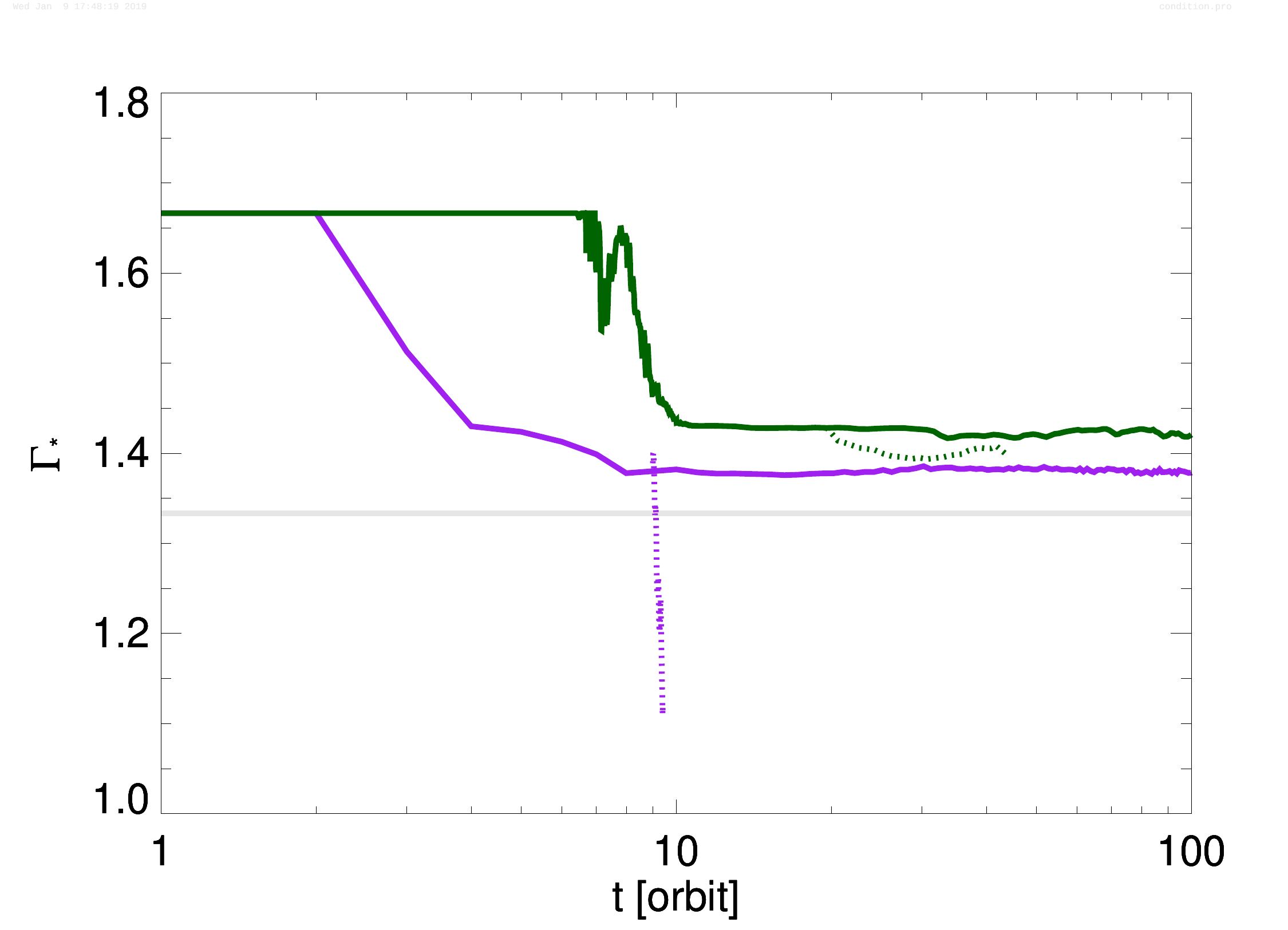} 
  \caption{Same as Fig. \ref{fig:condition_hikaku_gamma}, except for run R2 (green) and $\Sigma_0 = 60$ gcm$^{-2}$ at $r = 75$ AU (purple). The dotted curves correspond to double-resolution versions of the two cases, restarted from $t=9$ (purple) and $t=20$ (green) orbits, respectively.}
  \label{fig:condition_reso}
\end{figure}

Next we examine the spacial resolution. In Fig. \ref{fig:condition_reso}, we plot the time evolution of the adiabatic exponent $\Gamma_\sm$ for two cases; one is the case of $\Sigma = 30$ gcm$^{-2}$ at $r = 90$ AU (run R2; green) and the other is $\Sigma = 60$ gcm$^{-2}$ at $r = 75$ AU (purple). In the former case, a pressure-supported clump was formed and survived for many orbits, where $\Gamma_\sm\sim 1.42$ (thick green curve). In the latter case, a pressure-supported clump was formed similarly, but $\Gamma_\sm$ was closer to the critical value of $4/3$. To observe the dependence on the spacial resolution for these two cases, we doubled the number of cells, i.e. $(N_x,N_y,N_z) = (256, 256, 128)$, and restarted the calculation from a snapshot of the standard resolution run. The restarting time was set after a pressure-supported clump was formed. In the former case, the result did not change significantly although $\Gamma_\sm$ in the high-resolution run (thin green curve) was slightly lower, probably because mass concentration was resolved better. On the other hand, in the latter case, the result was changed drastically by doubling the spacial resolution. As shown by the thin purple curve, $\Gamma_\sm$ quickly decreased below the critical value of $4/3$ and runaway collapse occurred. This is because mass concentration enhanced by the doubled resolution was large enough to raise the core temperature above the H$_2$ dissociation temperature.

In summary, it is difficult to determine a precise condition for runaway collapse using local shearing box simulations with a fixed spacial resolution because whether runaway collapse occurs does depend on the box size and the spacial resolution. Global disk simulations are needed to determine the amount of mass involved in fragmentation, and a sort of mesh refinement is required to follow mass concentration by self-gravity at smaller scales. On the other hand, the fragmentation condition itself should be obtained by local shearing box simulations if the critical wavelength of GI is contained in the box and is resolved appropriately. 


\section{Summary}\label{sec:summary}

{\color{\modified}

  Using local three dimensional radiation hydrodynamics simulations, the nonlinear outcome of gravitational instability in an irradiated protoplanetary disc is investigated in a parameter space of the surface density $\Sigma$ and the radius $r$.

  Starting from laminar flow, axisymmetric self-gravitating density waves grow first. Their degree of self-gravitating becomes larger when $\Sigma$ is larger or the cooling time is shorter at larger radii. The density waves eventually collapse owing to the non-axisymmetric instability, which results in either fragmentation or gravito-turbulence after a transient phase.
The boundaries between the two are found at $r \sim 75$ AU as well as at $\Sigma$ that corresponds to the initial Toomre's parameter of $\sim 0.2$. The former boundary corresponds to the radius where the cooling time approaches unity. Even when the gravito-turbulence is established around the boundary radius, such short cooling time inevitably makes the fluctuation of $\Sigma$ large enough to trigger fragmentation. On the other hand, when $\Sigma$ is beyond the latter boundary (i.e. the initial Toomre's parameter is less than $\sim 0.2$), the initial laminar flow is so unstable against self-gravity that it evolves into fragmentation regardless of the radius or, equivalently, the cooling time. In other words, the initial gravitational energy is so large compared with the thermal energy that any heat generated in the nonlinear evolution of GI cannot compensate for it, and thus the gravito-turbulence of $Q \sim 1$ is not established. Runaway collapse follows fragmentation when the mass concentration at the centre of a bound object is high enough that the temperature exceeds the H$_2$ dissociation temperature.

  The fragmentation boundary found at $r \sim 75$ AU is consistent with a consensus in the literature in that the cooling time is essential for fragmentation \citep[e.g.][]{Gammie:2001}. On the other hand, another boundary found at $\Sigma \sim \Sigma_\text{0.2}$ indicates the importance of $Q_0$ \citep[c.f.][for global disc simulations]{Tsukamoto:2015,Takahashi:2016}, supporting the idea raised by \citet{Boss:2017} that the evolution of discs toward low $Q_0$ must be taken into account when assessing disc fragmentation possibilities.

  Also, we showed that the two fragmentation boundaries in our simulations are consistent with the linear analysis of the non-axisymmetric instability \citep{Takahashi:2016} when it is applied to the initial axisymmetric density waves. This indicates some connection between the local and global simulations of self-gravitating discs because fragmentation in global simulations is also explained by the linear analysis \citep{Takahashi:2016}.

  We have incorporated into our 3D simulations a realistic EOS, realistic radiative transfer (in the framework of FLD), and consider irradiation heating. These are relevant physics aspects for correct thermodynamic analysis related to protoplanetary discs. Actually, in Section \ref{sec:steady_accretion}, we showed that resolving the vertical structure with appropriate radiative transport is essential in determining the radial structure of the disc.
  However, there remain some limitations in our methods, and we add caveats here. Firstly, since we are using the local shearing box approximation, so our results should be valid in the case where the global transport of energy is not important, as discussed in detail in Paper I. Also, as we discussed in Section \ref{sec:dependence_on_numerical_issues}, the problem of whether runaway collapse occurs after fragmentation remains subtle, as the mass concentration at the centre of a formed clump is not properly solved in our simulation box with fixed size and resolution. Finally, we note that our study in this paper is dedicated to a particular protoplanetary disc system. Therefore, the fragmentation boundaries presented here may change if, for example, the central star's irradiation or the dust opacity is changed. 
}

\section*{Acknowledgements}
We thank Sanemichi Takahashi, Shu-ichiro Inutsuka and Yusuke Tsukamoto for valuable discussions, and Kengo Tomida for providing us with his EOS tables for our simulations. We also thank the anonymous referee for his/her valuable comments for improving the manuscriput. Numerical calculations were carried out partly on Cray XC30 at CfCA, National Astronomical Observatory of Japan, and partly on Cray XC40 at YITP in Kyoto University. SH was supported by Japan JSPS KAKENH 15K05040/18K03716 and the joint research project of ILE, Osaka University. JS was supported in part by the National Science Foundation under grant PHY-1144374, "A Max-Planck/Princeton Research Center for Plasma Physics" and grant PHY-0821899, "Center for Magnetic Self-Organization".




\bibliographystyle{mnras}
\bibliography{untitled} 




\appendix
\section{List of all runs}\label{sec:averaging_period}
\begin{figure*}
  \includegraphics[width=0.72\textwidth]{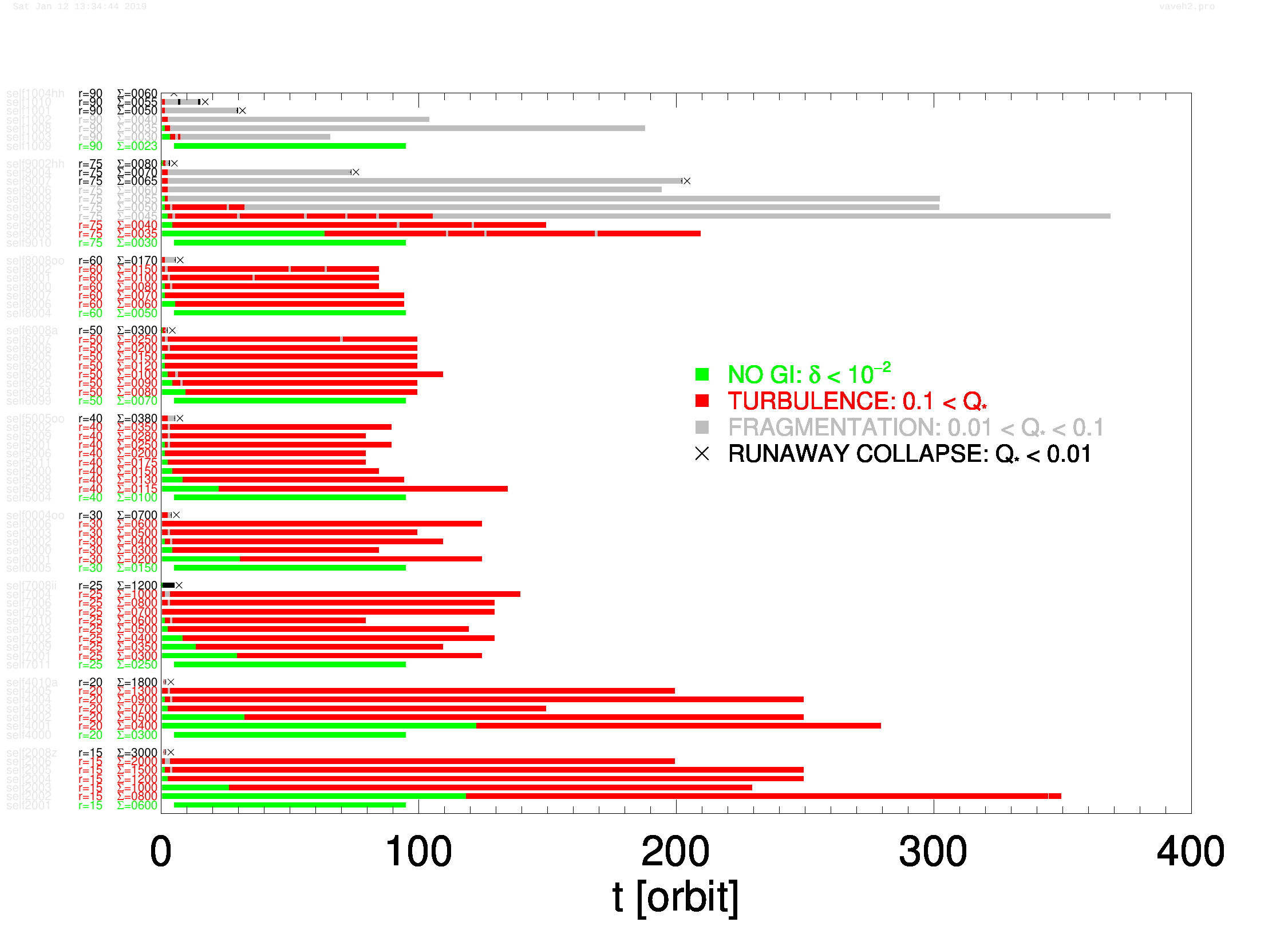}
  \caption{Running period for all runs. The two parameters, $r$ and $\Sigma_0$, of each run are shown on the left. (Job IDs are shown in the left-most column.) For runs where runaway collapse occurs, a cross is shown on the right of the period.}
  \label{fig:phase_diagram2}
\end{figure*}

Fig. \ref{fig:phase_diagram2} shows the running period of all {\numberOfSimulations} runs with the values of the two parameters, $r$ and $\Sigma$. The runs with a non-standard box size or resolution discussed in Section \ref{sec:dependence_on_numerical_issues} are excluded here. The running period is colour-coded with green, red, grey, and black for no GI, gravito-turbulence, fragmentation, and runaway collapse, respectively. As stated in the Section \ref{sec:nonlinear_development}, gravito-turbulence, fragmentation, and runaway collapse are distinguished by the value of $Q_\sm$. ``No GI'' is identified as a state where the fractional density fluctuation $\delta\Sigma/\Sigma$ defined in equation (\ref{eq:density_fluctuation}) is less than $10^{-2}$; such condition is adopted because the fluctuation $\delta\Sigma/\Sigma$ is not exactly zero even when GI does not occur, because the initial flow is not in an exact hydrostatic and thermal equilibrium. 
%


\bsp	
\label{lastpage}
\end{document}